\documentclass{article}

\usepackage[T1]{fontenc} % Modern font encoding
\usepackage{float}       % For creating charts, graphs and schemes
\usepackage{helvet}      % Helvetica font for sans serif
\usepackage{mathptmx}    % Times font ("Word-like")
\usepackage{etoolbox}
\usepackage{setspace}    % For double-spacing

\usepackage{bm}
\usepackage{graphicx}
\usepackage{amssymb}
\usepackage{amsmath}

\usepackage{caption}
\usepackage{fancyhdr}
\usepackage{float}
\usepackage{xcolor}

\usepackage{pdfpages}

\begin{document}

\centerline{\bf Origin of the Temperature-Induced Gap Bowing of Formamidinium-Methylammonium}

\centerline{\bf Lead Iodide Perovskites: Role of Cationic Rattlers}

\vspace{0.5cm}

%\end{document}

\centerline{Kai Xu$^{1}$, Adri\'{a}n Francisco-L\'{o}pez$^{1}$, Bethan L. Charles$^{2,3}$, M.~Isabel Alonso$^{1}$,}

\centerline{Miquel Garriga$^{1}$, Mark T.~Weller$^{2,4}$, and Alejandro R.~Go\~ni*$^{1,5}$}

\vspace{0.5cm}

\noindent
$^{1}$Institut de Ci\`encia de Materials de Barcelona, ICMAB-CSIC, Campus UAB, 08193 Bellaterra, Spain

\noindent
$^{2}$Dept. of Chemistry \& Centre for Sustainable Chemical Technologies, University of Bath, Claverton Down, Bath BA2 7AY, UK

\noindent
$^{3}$Dept. of Mechanical Engineering, Queens Building, University of Bristol, Bristol BS8 1TR, UK

\noindent
$^{4}$Dept. of Chemistry, Cardiff University, Wales CF10 3AT, UK

\noindent
$^{5}$ICREA, Passeig Llu\'is Companys 23, 08010 Barcelona, Spain
*Email: goni@icmab.es

\vspace{1cm}

\vspace{0.5cm}

\begin{abstract}

A thorough understanding of the temperature dependence of semiconductor band gaps is essential for optimizing optoelectronic devices. In this respect, the origin of the pronounced temperature-induced gap bowing observed in low-temperature phases of formamidinium-methylammonium (FA-MA) lead iodide perovskites has remained elusive until now. By combining temperature and pressure-dependent photoluminescence measurements on a series of FA$_x$MA$_{1-x}$PbI$_3$ mixed-cation single crystals with $x\in[0,1]$, we unravel the origin of this bowing. Both thermal expansion as well as electron-phonon interaction effects are responsible. However, the latter is the leading term, driven by the activation of an anomalous electron-phonon coupling mechanism linked to mixed vibrational modes, which combine inorganic-cage phonons involving octahedral tilting with low-frequency FA librations, i.e., FA \textit{rattler} modes. This occurs in the orthorhombic and (pseudo)tetragonal low-temperature phases, presumably featuring stripe domains with alternating octahedral tilt-axis patterns for FA concentrations between 20\% and 90\%. In this way, we have shed light on an intriguing behavior of lead halide perovskites that directly affects their optoelectronic properties.

\end{abstract}

\newpage

%\section{Introduction}

Semiconductors are characterized by an energy band gap typically located in the visible spectral region, from the near-infrared to the near-ultraviolet. Therefore, a fundamental understanding of what determines the semiconductor band gap and its temperature dependence is essential for the design and performance optimization of optoelectronic devices. For instance, the gap determines both the efficiency of a solar cell \cite{nrelx25a} and the color of a light-emitting device \cite{linxx18a}. Generally speaking, the effects of temperature on the band structure of any semiconductor are described by two terms related to thermal expansion (TE) and electron-phonon (EP) interaction \cite{laute85a,gopal87a,laute87a}. The former arises from the intrinsic anharmonicity of the crystal potential, which leads to expansion or contraction of the crystal lattice when temperature is raised or lowered, respectively. The gap thus partly changes due to the temperature-induced volume variations. It is important to note that the TE term can be experimentally assessed by determining the gap pressure coefficient in high-pressure experiments \cite{gonix98a}. In contrast, the energy renormalization of the electronic states produced by lattice vibrations is essentially due to the smearing of the crystal potential and the scattering of electrons by (virtual) phonons. Both effects are proportional to the Bose-Einstein phonon occupation number, hence the EP term \cite{cardo89a}. At a given temperature, the magnitude and sign of the EP term can be directly obtained from the difference of two experimentally determined quantities: the measured temperature slope of the gap and the TE term derived from the measured gap pressure coefficient \cite{rubin21a}. However, it is a common practice to describe the EP term using the Einstein-oscillator model \cite{goebe98a,serra02a,bhosa12a}, which considers an effective electron-phonon coupling constant for phonons with an average frequency derived from peaks in the phonon density of states. For a discussion of the current status and challenges in accounting for the electron-phonon renormalization in electronic structure calculations we refer to a recent perspective work \cite{shang23a}. 

Metal halide perovskites (MHPs) with formula AMX$_3$, with divalent Pb or Sn in the metal M-site, Cs, methylammonium (MA) or formamidinium (FA) as A-site cation, and where X is a halogen atom (Cl, Br or I), exhibit around room temperature a linear dependence on temperature of the gap with positive slope. This was obtained after comparing the near-ambient temperature gap slope of ca. twenty MHPs in nanocrystalline, thin film or bulk form (see Ref. \cite{fasah25a} and references therein). For MAPbI$_3$ it has been shown that this positive temperature gap slope picks up similarly strong contributions from thermal expansion and electron-phonon interaction effects \cite{franc19a}. However, a recent breakthrough in electron-phonon calculations in locally disordered (polymorphous) MHPs \cite{zacha23a} indicates that the relative weight of EP term can vary from 27\% to 97\% with increasing local disorder, going from FA, to MA and to Cs-containing perovskites \cite{zacha26a,zacha26b}. In any case, for the correct account of the gap temperature dependence the TE term cannot be neglected, even less in the low temperature phases, for which the gap pressure coefficient is strongly temperature dependent, as shown for MAPbI$_3$ \cite{pieni23a}, MAPbBr$_3$ \cite{pieni25a} and will be shown here for the FA$_x$MA$_{1-x}$PbI$_3$ system. The calculations within the polymorphous model reproduce very well the sign and magnitude of the linear temperature dependence observed experimentally at around room temperature and above for more than twelve MHP compounds \cite{zacha26b}. A notable exception to the \textit{linearity} rule is CsPbCl$_3$, which exhibits a sign reversal in the temperature slope, i.e. the gap decreases with increasing temperature near ambient, both for nanocrystals \cite{saran17a} and thin films \cite{xuxxx23a}. As recently reported elsewhere \cite{fasah25b}, we demonstrated that only the electron-phonon interaction is responsible for the sign reversal, which occurs due to the
activation of an anomalous electron-phonon coupling mechanism linked to vibrational modes characterized by synchronous octahedral tilting and the motion of the Cs cations inside the cage voids, the so-called Cs rattlers. These modes have been also invoked to understand the extremely low thermal conductivity of CsPbBr$_3$ \cite{lahns24a}. Another exception is the pronounced temperature-induced gap bowing exhibited by FA/MA mixed-cation lead iodide perovskites in low-temperature phases below ca. 250 K  \cite{zheng17a,franc20a}, the origin of which has remained elusive until now. In this letter we will show that the only way to account for this bowing is to introduce an additional Einstein oscillator in the EP term with negative amplitude, associated with an anomalous EP coupling term linked to FA-rattler modes instead of Cs. This occurs mainly in a low-temperature (pseudo)tetragonal phase, presumably featuring stripe domains with alternating (90 deg.) octahedral tilt-axis patterns for FA concentrations between 20\% and 90\% \cite{haine25a}.     

Given the importance for MHPs of the unique interaction between crystal structure and A-site cation dynamics for their electronic (in particular, the gap) and vibrational (phonon spectrum) properties, it will be instructive to review the composition-temperature phase diagram of the FA$_x$MA$_{1-x}$PbI$_3$ system to find a suitable explanation for what might be triggering the aforementioned anomalous EP coupling with the FA rattlers. For a general but complete account of the relationship between crystal structure and cation dynamics in MHPs we refer to the work of Simenas et al. \cite{simen24a}. Different parts of the FA$_x$MA$_{1-x}$PbI$_3$ phase diagram have been addressed by several authors \cite{weber16a,weber18a,mohan19a,franc20a,dutta25a,haine25a}. The relationship between structure and A-site cation dynamics has been revisited from two antagonist points of view regarding the interaction mechanism, that is hydrogen bonding \cite{konto20a,weixx23a} and dynamic steric interaction \cite{xuxxx23b}. The influence of the latter has been also considered to understand phonon interactions in MHPs as revealed by Raman scattering \cite{gonix26a}. Of particular interest for this work is the very recent molecular dynamics (MD) study of the phase diagram of FA$_x$MA$_{1-x}$PbI$_3$, linking structure, dynamics and electronic band structure \cite{haine25a}. The main result is the identification of a morphotropic phase boundary (MPB) for FA contents of ca. 27\%, delineating the transition from the tetragonal phase with I4/mcm space group \cite{welle15a} and (in Glazer notation) out-of-phase $a^0a^0c^-$ octahedral tilt pattern, characteristic of the MA-rich lead iodide compounds, to the also tetragonal phase with P4/mbm symmetry \cite{weber16a,weber18a} but in-phase  $a^0a^0c^+$ tilt pattern, typical of solid solutions with high FA content. Interestingly, this transition is accompanied by the formation of stripe domains structures with alternating tilt-axis patterns and, most importantly, the enhancement of dynamic disorder and electron-phonon coupling \cite{haine25a}. Here we would like to propose that the (pseudo)tetragonal phase stabilized below approx. 250 K and at intermediate FA concentrations, called the Tetra-III phase \cite{franc20a}, has a layered character exhibiting a mosaic-like structure, where nanoscale regions with an alternating tilt pattern coexist within each stripe domain. In fact, for this phase the space group assignment from neutron scattering \cite{weber18a} is uncertain, being P4bm only plausible, but indicating a certain degree of structural disorder present. This is the phase for which mainly, but not exclusively, the gap temperature dependence displays the pronounced bowing.     

\begin{figure}[H]
\includegraphics[width=13cm]{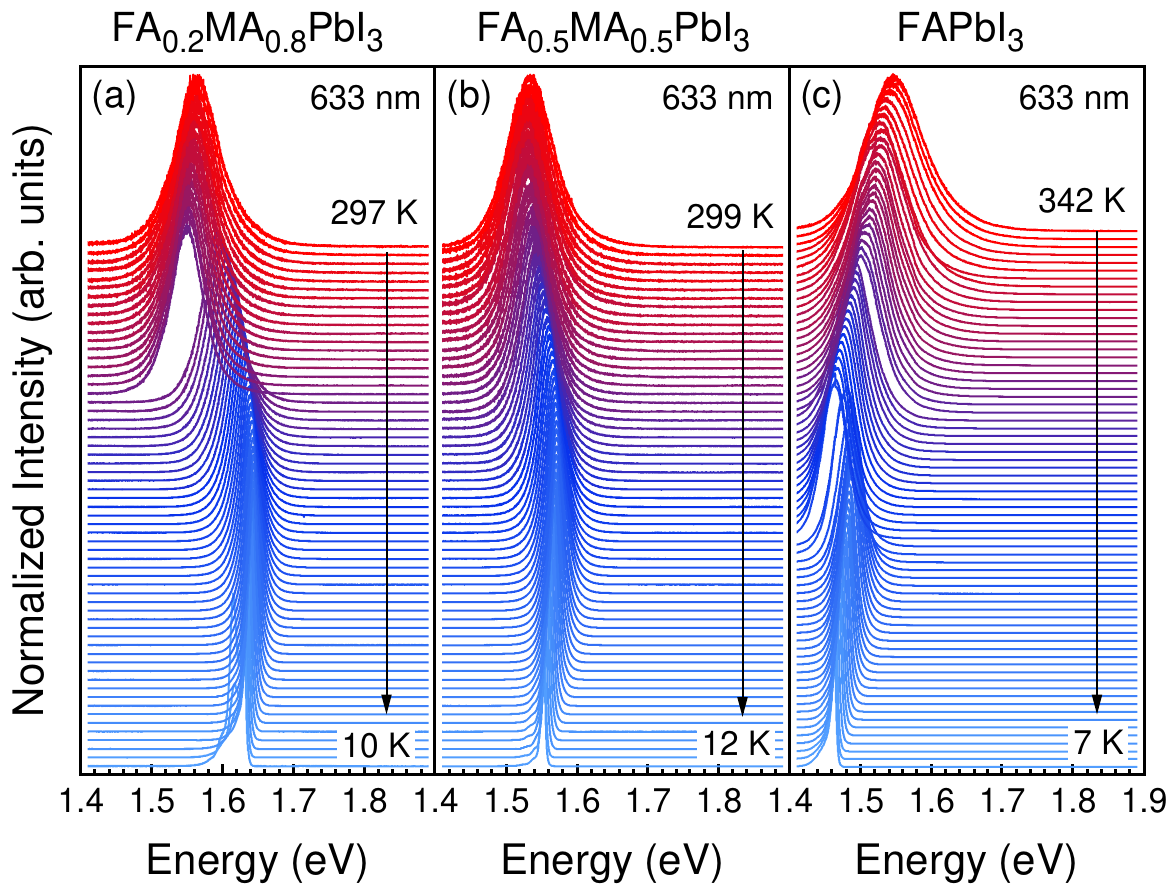}
\caption{
\label{PL spectra}
PL spectra of FA$_x$MA$_{1-x}$PbI$_3$ single crystals for three selected compositions, namely, (a) x = 0.2, (b) x = 0.5, and (c) x = 1.0, recorded at different temperatures using the red line (633 nm) for excitation. The spectra were normalized to their maximum intensities and plotted with a vertical shift for increasing temperature. The temperature range is indicated (a temperature step of ca. 5 K).
}
\end{figure}

%\vspace{0.5cm}

%\section{Experimental}
%\subsection{Synthesis and characterization of the mixed-halide NCs}

High-quality single crystals of FA$_{x}$MA$_{1-x}$PbI$_{3}$ across the full composition range were synthesized using the inverse solubility method of Saidaminov \textit{et al.} \cite{saida15a}, as explained in Note $\sharp$0 of the Supporting Information (S.I.). The photoluminescence (PL) spectra were excited with the 633 nm line of a He-Ne laser using a very low incident light power below 2 $\mu$W (power density <15 W/cm$^2$) to avoid any photo-degradation of the samples \cite{leguy16a}. PL spectra were corrected for the spectral response of the spectrometer by normalizing each spectrum using the detector and the 600-grooves/mm grating characteristics. Temperature-dependent PL measurements were carried out by decreasing the temperature between ca. 365 and 6 K in steps of 5 K, using liquid helium in a gas flow cryostat from CryoVac. Room temperature high-pressure PL measurements were performed employing a gasketed diamond anvil cell (DAC). Anhydrous propanol was used as pressure transmitting medium at room temperature, which ensures good hydrostatic conditions and proved chemically inert to (FA,MA)PbI$_3$. In contrast, for the high-pressure PL measurements at low temperatures down to 4 K a specially designed He-bath cryostat (Konti-IT from Cryovac) was used that can allocate the DAC in its cold bore \cite{gonix98a}. For cryogenic high-pressure measurements, helium is the best possible pressure medium. The difficulty is that loading ought to be performed with the DAC immersed in superfluid helium, as explained in detail in the Note $\sharp$1 of the S.I.

%\section{Results}

\begin{figure}[H]
\includegraphics[width=12cm]{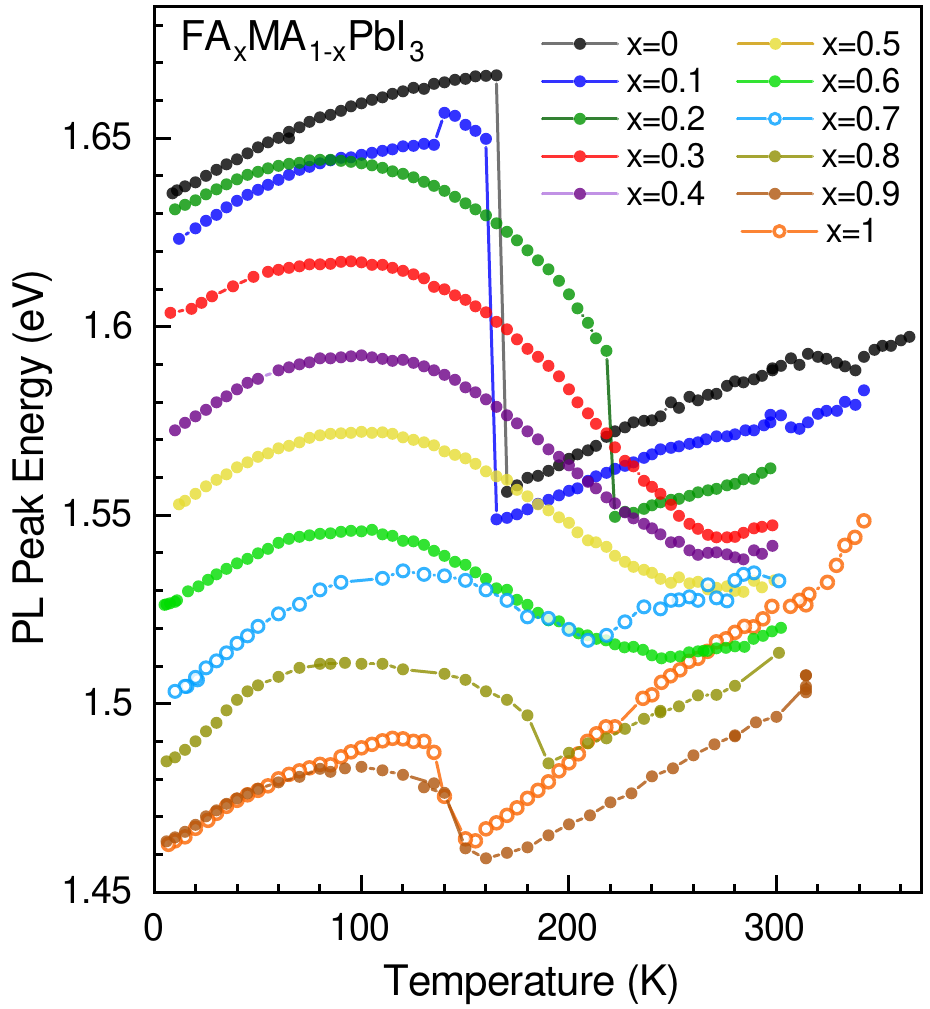}
\caption{
\label{Eg-vs-T}
Maximum PL peak energy position plotted as a function of temperature, obtained from the PL line shape fits using a cross-product function (Eq. 1 of the S.I.) for the complete set of compositions of the
FA$_x$MA$_{1-x}$PbI$_3$ series.
}
\end{figure}

Figures \ref{PL spectra}a-c display the temperature evolution from ca. 6 to 365 K of the PL spectra of FA$_x$MA$_{1-x}$PbI$_3$ single crystals for three selected compositions, namely, $x$=0.2, 0.5, and 1.0, respectively. The spectra cascade for the rest of FA compositions were published elsewhere \cite{franc20a}. All spectra were normalized to its absolute maximum intensity and vertically offset for clarity. To analyze the PL spectra, we used a Gaussian-Lorentzian cross-product function for describing the main peak which is ascribed to free-exciton recombination \cite{galco17a}. The procedure for the line-shape analysis of the PL spectra is described in Note $\sharp2$ of the S.I. 
The values of the PL peak maximum obtained from a line-shape analysis of the PL spectra fits are plotted as a function of temperature in Fig. \ref{Eg-vs-T} for the eleven compositions of the FA$_x$MA$_{1-x}$PbI$_3$ series.
For practical purposes we consider the shift of the PL peak energy with temperature to be representative of the shift of the gap \cite{rubin21a,hanse24a}. The sets of data points exhibit a clear decreasing trend of the gap at room temperature with increasing FA content, except for the two compositions $x$=0.7 and 1 (open symbols in Fig.  \ref{Eg-vs-T}). This might be due to a poorer quality of these particular single crystals. In the case of 70\% FA content, a spectroscopic ellipsometry study \cite{alons19a} indicates that this sample consists of FA and MA-rich segregated regions separated by areas with high concentration of vacancies (predominantly of MA \cite{orans18a,meggi18a,franc21a}), although maintaining an average stoichiometry of $x=0.7$. For FAPbI$_3$, in contrast, we were not able to study the pristine as-grown sample. For technical reasons, the sample had already transformed into the yellow hexagonal $\delta$-phase when received. Although we were able to transform the FAPbI$_3$ crystal back to the perovskite $\alpha$-phase by heating it on a hotplate to about 80 $^\circ$C for a few minutes \cite{charl17a}, it seems that this treatment was somewhat detrimental for its crystal quality.  

Here we aim to elucidate the pronounced temperature-induced gap bowing in the low-temperature phases of FA$_x$MA$_{1-x}$PbI$_3$ solid solutions, as displayed in Fig. \ref{Eg-vs-T} and also observed elsewhere \cite{zheng17a}. We realized in a previous study on MAPbI$_3$ \cite{franc19a} that it is more convenient for subsequent analysis to employ the first derivative of the gap with respect to temperature. For this purpose we derived numerically, point-by-point, the previously smoothed gap-vs-temperature curves of Fig. \ref{Eg-vs-T}. The results for all compositions are shown in Figs. S0a-k on Note $\sharp$3 of the S.I. It is worth noticing, first, that in the low-temperature phases, the first derivatives are all pretty linear over a large temperature range of more than 100 K, indicating that the bowing is an almost perfect parabola. Second, the only first derivatives that do change sign, becoming negative in some temperature interval, are the ones for FA contents $0.2\leq x\leq0.9$. Thus, it seems that a certain amount of disorder due to A-site cation mixing is necessary for having a sign turnover in the gap temperature dependence. 
To understand such a behavior we have first to disentangle the effects of thermal expansion and electron-phonon interaction on the gap temperature renormalization \cite{franc19a,rubin21a,perez23a}. According to Eq. (2) in Note $\sharp$4 of the S.I., the derivative of the gap over temperature contains solely the thermal expansion (TE) term and the one due to electron-phonon interaction (EP) \cite{laute85a,gopal87a,goebe98a}.
The effect on the gap due to the lattice contraction with decreasing temperature is intimately related to the response of the electronic states under hydrostatic pressure \cite{laute85a,gopal87a}:
\begin{equation}
\left[\frac{\partial E_g}{\partial T}\right]_{TE}=-\alpha_V\cdot B_0\cdot\frac{dE_g}{dP},
\label{TE}
\end{equation}
\noindent where $\alpha_V$ is the volumetric thermal expansion coefficient, $B_0$ is the bulk modulus and $\frac{dE_g}{dP}$ is the pressure coefficient of the gap, determined here from high pressure experiments. 

From the experimental point of view, the main achievement of this work is the determination of the gap pressure coefficient from PL spectra recorded at different pressures for a set of temperatures different (lower) than ambient. Since these experiments are very challenging and time consuming, we were able to perform a restricted number of measurements at a few but selected temperatures in the range of the bowing for seven representative FA compositions. The main difficulty lies in the fact that the measurements must be performed within a very small pressure range up to 0.5-0.7 GPa, which, based on experience, is the range in which the first pressure-induced phase transition occurs \cite{franc18a,posto17a}, and this must be avoided. For subsequent measurements at a different temperature, the pressure must be reduced for the next upstroke, always at room temperature. During the process of reducing the temperature to the new desired value, the DAC often opens unintentionally, releasing the pressure medium and requiring a new helium (cryogenic) reload. The full set of results regarding the gap values versus pressure, as extracted from the PL spectra, are displayed for the selected temperatures in Figs. S1-S7 in Note $\sharp$4 of the S.I. for $x$=0, 0.2, 0.4, 0.5, 0.6, 0.8, and 0.9, respectively. The slopes of the linear fits to the data points (red solid lines) directly give the gap pressure coefficient which are listed in the corresponding S.I. Tables S1 to S7. In good quantitative and qualitative agreement with previous observations for MAPbI$_3$ \cite{pieni23a} and MAPbBr$_3$ \cite{pieni25a}, for all studied FA compositions, the sign and magnitude of the gap pressure coefficient $\frac{dE_g}{dP}$ are strongly dependent on temperature, as is the TE term as well.

To calculate the TE term according to Eq. (\ref{TE}), we need the values of the volumetric thermal expansion coefficient $\alpha_V$ as a function of temperature, i.e., for the different structural phases adopted by the material, and for all studied compositions. Unfortunately, this kind of data are only available for MAPbI$_3$ \cite{jacob15a,whitf17a,whitf16a}, FA$_{0.5}$MA$_{0.5}$PbI$_3$ \cite{weber16a} and FAPbI$_3$ \cite{fabin16a}. Regarding the bulk modulus $B_0$, the available information is similarly scarce; in fact, it is solely available for MAPbI$_3$ \cite{jaffe16a,szafr16a} and FAPbI$_3$ \cite{corde19a} and only at room temperature. However, the evidence is that $B_0$ depends principally on the crystal structure, being temperature independent within the stability range of each phase \cite{jaffe16a,corde19a}. In short, we have been forced to make certain assumptions to compensate for the lack of information regarding  $\alpha_V$ and $B_0$, as explained in details in Note $\sharp$4 of the S.I. In Tables S1 to S7 we have listed all the information needed to compute the TE term for the different compositions, temperatures, and crystal phases measured. Strikingly, strong variations in magnitude and even a change in sign of the TE term is observed at intermediate FA concentrations for the low temperature (Ortho \& Tetra-III) phases. However, we show below that these changes in the TE term only partially explain the temperature-induced bowing; the main contribution comes from changes in the electron-phonon interaction.

Regarding the contributions to the gap temperature renormalization stemming from electron-phonon interactions, the most important ones arise from peaks in the phonon density of states (DOS) \cite{gopal87a}. This is at the origin of the Einstein-oscillator model \cite{goebe98a,serra02a,bhosa12a}, which approximates the different contributions to the EP term by oscillators with effective amplitude $A_i$ and phonon eigen-frequency $\omega_i$, the latter inferred from the peaks in the phonon DOS (see Eq. (6) in Note $\sharp$5 of the S.I.). The EP correction to the gap is obtained by calculating analytically the first derivative over temperature of the Bose-Einstein occupation factor $n_B(\omega_i,T)=\left(e^{\beta\hslash\omega_{i}}-1\right)^{-1}$, with $\beta=\frac{1}{k_BT}$:

\begin{equation}
%\begin{split}
\left[\frac{\partial E_g}{\partial T}\right]_{EP} = \sum_i \frac{A_i}{4T}\cdot\frac{\hslash\omega_i}{k_BT}\cdot\frac{1}{sinh^2\left(\frac{\hslash\omega_i}{2k_BT}\right)}. %\sum_i A_i\cdot\frac{\partial\left(n_B(\omega_i,T)+\frac{1}{2}\right)}{\partial T} \\
%                                                  & =
%\end{split}
\label{EP}
\end{equation}
\noindent 

\begin{figure}[H]
\includegraphics[width=6.5cm]{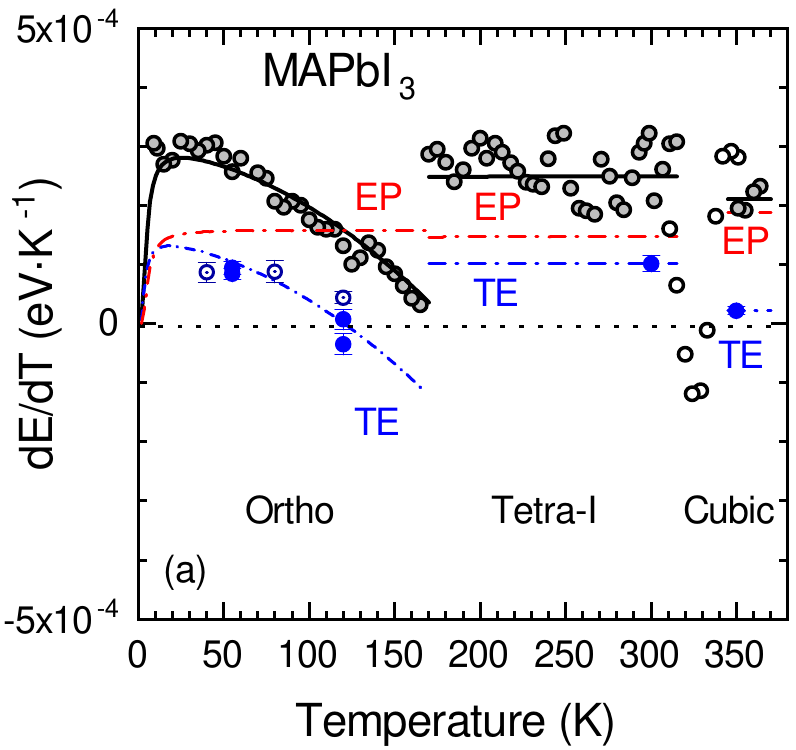}
\includegraphics[width=6.5cm]{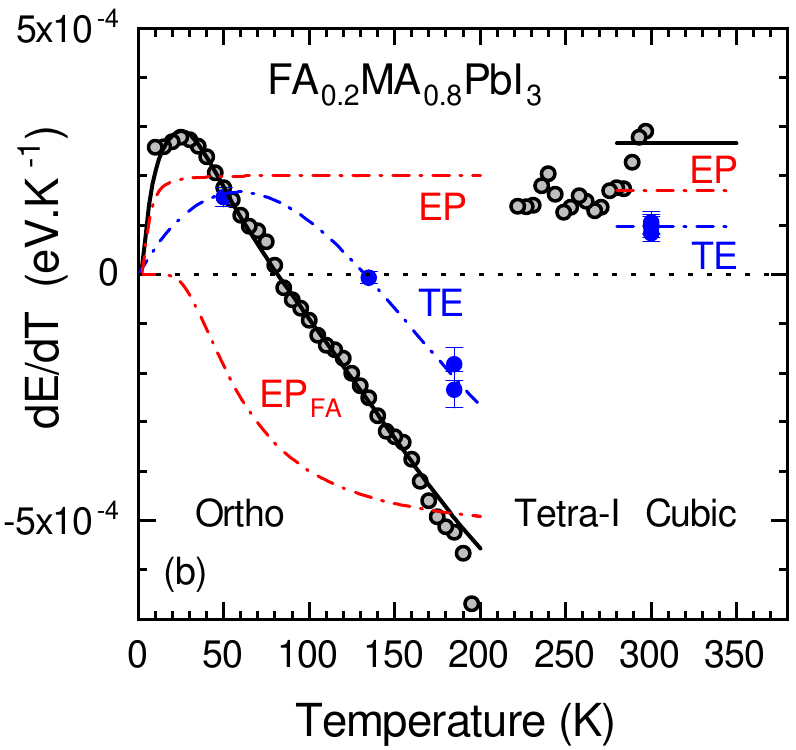}
\includegraphics[width=6.5cm]{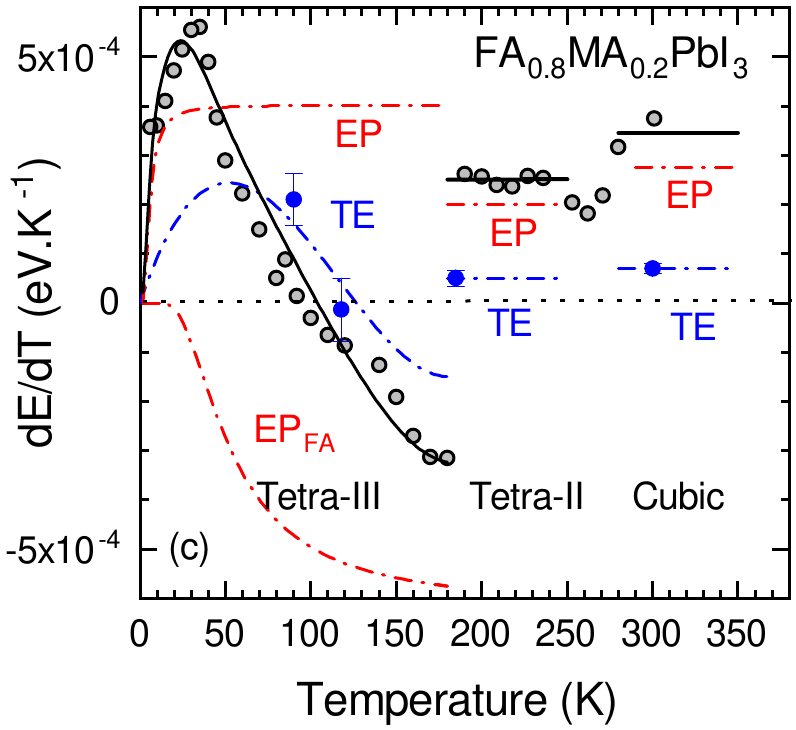}
\caption{
\label{dE-dT}
The first derivative of the gap energy with respect to temperature (black-grey symbols), numerically calculated from the smoothed data of Fig. \ref{Eg-vs-T} for FA$_x$MA$_{1-x}$PbI$_3$ mixed-cation crystals with (a) $x$=0, (b) $x$=0.2 and (c) $x$=0.8. The solid black lines represent a fit to \emph{only} the closed data points in the temperature range of stability of different phases, using the sum of thermal expansion (TE, blue dot-dashed curve and symbol) and electron-phonon interaction (EP, red dot-dashed curve). In (a), open blue symbols correspond to the values of the TE term calculated using the gap pressure coefficients determined for MAPbI$_3$ at low temperatures \cite{pieni23a}. In (b) and (c), the curve labeled EP$_{FA}$ represents the additional Einstein oscillator introduced to account for the anomalous FA-rattler related EP coupling. See text for details.
}
\end{figure}

To evaluate the EP term we thus have to perform a least-squares fit to the previously computed first-derivative data points using the sum of the TE and EP terms calculated according to Eq. (\ref{TE}) and Eq. (\ref{EP}), respectively. The black-grey symbols in Figs. \ref{dE-dT}a-c correspond to the first derivative data sets for a FA content of $x$=0, 0.2 and 0.8; three representative compositions regarding the temperature-induced gap bowing (the results for the seven measured compositions are shown in  Figs. S8 to S14 in Note $\sharp$5 of the S.I.). To obtain the EP term (or terms) for each composition by fitting, the analysis of the first-derivatives was performed piecewise within the temperature range of stability of the different phases, as indicated in the figures. The blue symbols and the blue dot-dashed curves in Figs. \ref{dE-dT}a-c correspond to the TE term resulting from Eq. (\ref{TE}) and their polynomial fits, respectively, using the parameters tabulated in Tables S1 to S7 on Note $\sharp$4 of the S.I. At high temperatures, i.e. for the Cubic and Tetra-I\&II phases, the TE term is temperature independent, as is common in MHPs \cite{fasah25a,zacha26b}. The analysis of the gap temperature dependence for these phases can be found in Note $\sharp$5 of the S.I. In the following, we will restrict the discussion exclusively to the behavior of the gap in the low-temperature phases, that is the Ortho and Tetra-III phases, for which the bowing is usually observed. 

\begin{table}[H]
\caption{Einstein-oscillator parameters corresponding to the effective amplitude $A_i$ and frequency $\omega_i$ describing the contribution from electron-phonon interaction $\left[\frac{\partial E_g}{\partial T}\right]_{EP}$ to the renormalization of the band gap for the indicated crystal phase of FA$_{x}$MA$_{1-x}$PbI$_{3}$ single crystals with the seven FA concentrations $x$ measured. The numbers in parentheses represent the error bars (uncertainty of the last digits) of the adjustable fitting parameters; conversely, numbers without error bars mean that this parameter was kept fixed during fitting. Also indicated is the temperature range in which the fit is excellent.}
\vspace{0.5 cm}
\begin{tabular}{| c | c | c c c | c |}
\hline
% & \multicolumn{4}{c}{\textbf{FA$_{0.5}$MA$_{0.5}$PbI$_{3}$}} & \\
% \hline
% & & & & &\\
 $x$  &  Phase    &~Einstein  ~&~  $A_i$  ~~&$\hslash\omega_i$& $T$ fitting range \\
      &           & oscillator &   (meV)    &     (meV)       &      (K)          \\
\hline
\hline
% & & & & &\\
 0    & Ortho     & normal     &  2.7(5)    &     1.5(10)     &       <165        \\
\hline
% & & & & &\\
      &           & normal     &  3.5(5)    &       1.5       &                   \\
 0.2  &Ortho      &            &            &                 &       <170        \\
      &           & FA rattler &  -98(10)   &      16(1)      &                   \\
\hline
% & & & & &\\
      &           & normal     &    6(1)    &       1.5       &                   \\
 0.4  &Tetra-III  &            &            &                 &       <130        \\
      &           & FA rattler &  -85(10)   &      14(3)      &                   \\
\hline
% & & & & &\\
      &           & normal     &    7(1)    &       1.5       &                   \\
 0.5  &Tetra-III  &            &            &                 &       <120        \\
      &           & FA rattler & -135(10)   &      18(2)      &                   \\
\hline
% & & & & &\\
      &           & normal     &    6(1)    &       1.5       &                   \\
 0.6  &Tetra-III  &            &            &                 &       <125        \\
      &           & FA rattler & -125(10)   &      18(3)      &                   \\
\hline
% & & & & &\\
      &Tetra-III  & normal     &    7(1)    &       1.5       &                   \\
 0.8  &    or     &            &            &                 &       <125        \\
      &Ortho      & FA rattler & -105(10)   &      14(2)      &                   \\
\hline
% & & & & &\\ 
      &Tetra-III  & normal     &    5(1)    &       1.5       &                   \\
 0.9  &    or     &            &            &                 &       <110        \\
      &Ortho      & FA rattler &  -75(10)   &      15(2)      &                   \\
\hline
\hline
\end{tabular}
\vspace{0.5 cm}
\label{EP all x}
\end{table}

An inspection of the gap first-derivative curves shown in Figs. S0a-k of the S.I. indicates that a turnover in the sign of the first derivative happens only for intermediate FA concentrations but not at the compositional ends ($x$=0, 0.1, and 1). Let us thus start discussing MAPbI$_3$ as representative of the latter case; the corresponding data are shown in Fig. \ref{dE-dT}a. The contribution from electron-phonon interaction is calculated using the function of Eq. (\ref{EP}) and, like for the room-temperature Tetra-I phase \cite{franc19a}, considering a single Einstein oscillator, being its amplitude and frequency the only adjustable parameters. The solid black curve in Fig. \ref{dE-dT}a represents the result of a least-squares fit to the data points of the Ortho phase of MAPbI$_3$ using the polynomial function describing the TE term (blue dot-dashed curve) plus the single Einstein-oscillator function (dot-dashed red curve). The resulting amplitude and frequency are listed in Table \ref{EP all x}. This oscillator has also a positive amplitude but a smaller frequency of 1.5 meV compared to that of the room-temperature oscillator (6 meV), as dictated by the lower occupation of higher energy phonons at low temperatures. This frequency coincides fairly well with the first peak in the phonon DOS of MAPbI$_3$ \cite{leguy16a}, corresponding to zone-edge acoustical phonon modes of the inorganic cage. It thus represents the \textit{normal} EP coupling term \cite{franc19a,fasah25a}. Interestingly, the non-linearity in the gap first derivative of the Ortho phase is entirely given by the temperature-dependent TE term (see Fig. \ref{dE-dT}a).

For intermediate FA concentrations the situation with the EP term changes dramatically. As can be seen in Figs. \ref{dE-dT}b,c for $x$=0.2 and 0.8, respectively (see Figs. S10, S11, S12, and S14 of the S.I. for the rest of compositions), at temperatures lower than around 100 K, the gap first derivative is always positive, as well as both the TE term and the normal EP term. Although within the stability range of the Ortho and Tetra-III phases the contribution from the TE term can become negative above 100 K, its contribution is clearly insufficient to account for the strongly negative values of the (total) temperature gap derivative. Hence, the only way to correctly describe the gap temperature dependence is to introduce an additional Einstein oscillator with negative amplitude. For the fits, we thus used for the EP term two Einstein oscillators: One with positive amplitude and a fixed frequency of 1.5 meV, as for MAPbI$_3$, to account for the normal EP coupling, and one with adjustable (negative) amplitude and frequency, called EP$_{FA}$, for it will be assigned to FA rattler modes, as previously done for Cs in CsPbCl$_3$ NCs \cite{fasah25b}. %fixed to a value of 4 meV, the frequency of the Cs rattle modes \cite{lahns24a} (see discussion below). This is necessary because, otherwise, the problem has multi-valued solutions. %, with large sets of $A_i$ and $\omega_i$ pairs that yield almost exactly the same fit quality. Nevertheless, the conclusions drawn do not depend on the exact values of the oscillator frequencies.
The red dot-dashed curves labeled EP$_{FA}$ in Figs. \ref{dE-dT}b,c and Figs. S8-S14 of the S.I. represent the contribution of the additional Einstein oscillator. The results of the least-squares fits for the seven compositions studied but concerning only the EP terms are shown in Table \ref{EP all x} and plotted as a function of FA composition in Figs. S15a,b of the S.I. We note that the gap first derivatives are perfectly described by the fitting function (black solid curves) within the temperature range indicated in Table \ref{EP all x}. Strikingly, the average frequency of the additional oscillator is ($16\pm2$) meV, in excellent agreement with the lowest-frequency FA libration (135 cm$^{-1}$=17 meV) reported for FAPbI$_3$ \cite{konto20a} or the average frequency of a mode with predominantly FA contribution and a fairly flat dispersion, indicating its localized nature, calculated for FAPbBr$_3$ \cite{zacha26b}. The necessary admixture of this FA-mode with the inorganic cage phonons can be observed in the piecewise finite dispersion and by the color variation of the dispersion curves shown in Fig. 14c of Ref. [\cite{zacha26b}], indicating a contribution from cage atoms, as expected for FA rattlers.  
The average amplitude of the additional Einstein oscillators is ($-105\pm25$) meV, thus being the leading term that overcompensates the effects of thermal expansion and the normal EP term. All these facts and the clear similarity with the phenomenology exhibited by the Cs rattlers in Cs$_x$MA$_{1-x}$PbI$_3$ single crystals (x = 0.05 and 0.1) \cite{perez23a} and CsPb(Br$_{1-x}$Cl$_x$)Cl$_3$ NCs with $x>0.5$ \cite{fasah25b} led us to the conclusion that the additional Einstein oscillator corresponds to FA-rattler modes which lead to the anomalous electron-phonon coupling that is the main cause of the observed gap bowing. 

%\section{Discussion}

Recently, band-structure calculations of the temperature dependence of band gaps in disordered solids have advanced considerably thanks to the development of a non-perturbative approach based on the idea of searching for the polymorphic structure that minimizes total energy, taking into account the thermal occupation of the lattice phonons \cite{zacha23a,zacha26a,zacha26b,zhaox20a}. In MHPs, polymorphism arises from dynamic disorder induced by the A-site cation dynamics \cite{evenx16a}. In brief, positional polymorphism represents multiple domains of local disorder at static-equilibrium that define a minimum in the system anharmonic potential energy surface (PES). Despite the presence of positional polymorphism, on a macroscopic level, the overall structure retains the average high-symmetry of the tetragonal and cubic phases. For MHPs the simple monomorphous network typically utilized by ab-initio electronic structure calculations is dynamically unstable, thus corresponding to a \textit{local} minimum on the PES. For the exploration of the polymorphous structures a unified treatment was developed \cite{zacha23a}, which relies on the anharmonic special displacement method \cite{zacha20a}. This is an efficient supercell approach, where atoms are displaced according to a special linear combination of the computed phonon eigenvectors with amplitudes determined by the associated root mean squared displacements. It allows for the generation of a single optimal structure that best represents the system at thermal equilibrium. For the (high-temperature) tetragonal and cubic phases of a variety of twelve Pb and Sn halide perovskite compounds, it is found by averaging over ten polymorphous structures that the gap increases nearly linearly with temperature with an average slope of 4$\times10^{-4}$ eV/K \cite{zacha26a,zacha26b}, often in excellent agreement with the experiment for materials featuring \textit{normal} EP coupling (see literature survey in Ref. [\cite{fasah25a}] and references therein). The question that immediately arises is how can a "static" band structure calculation account so well for an intrinsically "dynamic" property like the electron-phonon interaction and the temperature renormalization of the gap? The answer is simply found in the Born-Oppenheimer approximation which states that the lattice atoms are static for the crystal electrons, which move at speeds three to four orders of magnitude faster, just due to the extreme mass differences. Phonons just cause a local distortion of the lattice, which for crystal electrons does not vary in time but spatially. Statistically, the variations of the distortion of a given unit cell in time can be isomorphically mapped into the distortions of all other unit cells at a given time (snapshots of the phonon-induced deformations). Electrons just sample this polymorphous, spatially inhomogeneous structure, which is exactly the outcome of the polymorphous method. Regrettably, in the case of CsPbCl$_3$, for which the gap temperature slope at ambient was shown to be dominated by the effects of anomalous EP coupling \cite{fasah25b}, the polymorphous approach delivers a small though still positive slope \cite{zacha26b}. This might be an indication that rattler modes are not being effectively incorporated into the polymorphous methodology.  

\begin{figure}[H]
\includegraphics[width=12cm]{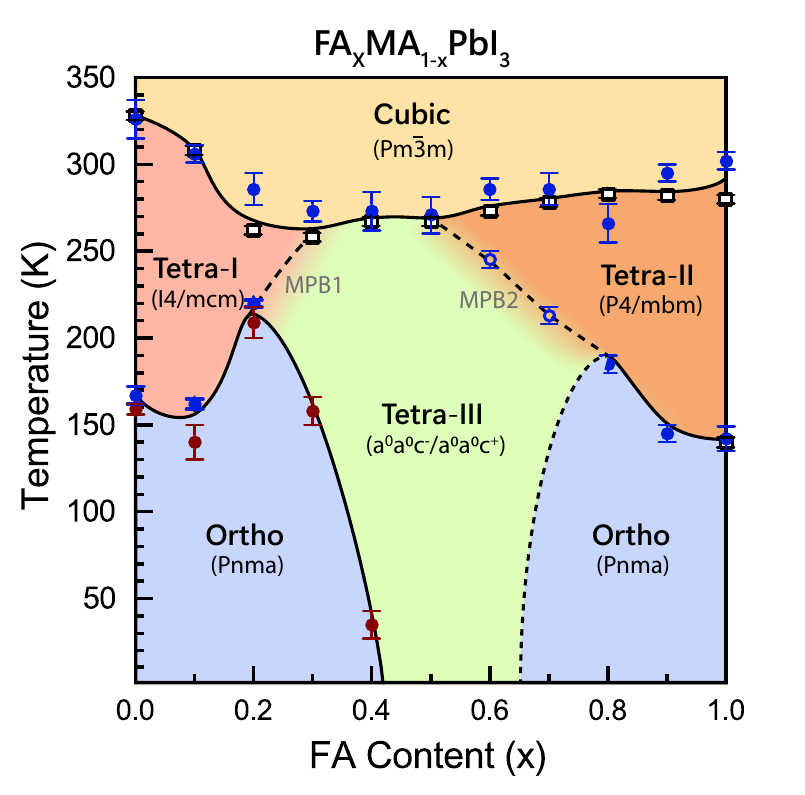}
\caption{
\label{Diagram}
Revised version of the temperature-versus-composition phase diagram of FA$_x$MA$_{1-x}$PbI$_3$ perovskite solid solutions \cite{franc20a}. The symbols represent experimental data (Raman and PL data from Ref. \cite{franc20a} and X-ray/neutron scattering data from Refs. \cite{welle15a,welle15b,weber16a,weber18a}). The different crystal structures and their symmetry are indicated. Phase separation lines as well as morphotropic phase boundaries (MPBs), indicated by dashed lines, are a guide to the eye.
}
\end{figure}

What is also well explained within the polymorphic method is the significant reduction of the band gap as MA is gradually replaced by FA (see Fig. \ref{Eg-vs-T}); its explanation remained elusive until now. Although the molecular orbitals of the organic cations play no role in defining the band-edge states, the A-site cation size and polarity strongly affect the amount of local disorder (particularly octahedral tilting) present in the polymorphous phases. The larger the cation size, the less disorder is locally generated, which leads to a smaller (positive) gap renormalization. The fact that the gap strongly opens with octahedral tilting is directly inferred from the observed jumps in the gap at the phase transition from Tetra-I/Tetra-II to Ortho (see Fig. \ref{Eg-vs-T} and Ref. [\cite{kongx16a}]). The computed gap opening within the polymorphous model is 0.27, 0.46 and 0.61 eV in FA-, MA-, and Cs-based compounds, respectively \cite{zacha26b}. This implies a gap difference between MAPbI$_3$ and FAPbI$_3$ of 190 meV, in very good agreement with experiment (see Fig. S15c on Note $\sharp$5 of the S.I.). To be fair, a similar composition dependence of the gap has been previously reported for the FA$_{x}$MA$_{1-x}$PbI$_{3}$ system, as computed by first-principles density-functional theory (DFT) calculations for structures where the orientations of the organic cations resulted from careful examination of the energy landscape \cite{senno21a}. These (zero-temperature) calculations indicate that H bonding is crucial for the stabilization of those crystal structures, which exhibit an increment in Pb-I-Pb bond angle from ca. 160 deg. to 175 deg., as FA content varies from 0 to 1 \cite{senno21a}. Concomitantly with this decrease in average octahedral tilting, leading to a more symmetric structure, the gap decreases linearly with increasing FA content, being the gap difference between the two end compositions ca. 300 meV, when spin-orbit coupling is considered.

To better understand the conditions leading to anomalous EP coupling, we will now revisit the composition-temperature phase diagram of FA$_{x}$MA$_{1-x}$PbI$_{3}$ in light of new theoretical developments. This phase diagram was recently mapped out using a machine-learned interatomic potential in molecular dynamics (MD) simulations \cite{haine25a}. The main result was the identification of a morphotropic phase boundary (labeled MPB1 in Fig. \ref{Diagram}) at around 27\% FA content, which separates the two tetragonal phases, named Tetra-I and Tetra-II, characteristic of both compositional end materials: MAPbI$_3$ and FAPbI$_3$, respectively. A key observation is that the Tetra-I and Tetra-II phases exhibit opposing tilt patterns (in Glazer notation, out-of-phase $a^0a^0c^-$ vs in-phase $a^0a^0c^+$); patterns which coincide with the Brillouin-zone edge R$_z$ and M$_z$ phonon mode eigenvectors ($z$ is the long tetragonal axis), respectively. The transition from one phase to another is not abrupt but gradual and coincides with a mode crossover at which R and M phonons become nearly degenerate \cite{haine25a}. An analysis of the evolution of the phonon mode projections indicate that, across the transition, there is a certain range of compositions around the MPB, where the material breaks up into a mosaic-like arrangement of layered regions in which either R$_z$ or M$_z$ modes are predominantly activated. 
Interestingly, from the point of view of the tilt patterns, the FA$_{x}$MA$_{1-x}$PbI$_{3}$ solid solutions behave as an eutectic system \cite{bergm92a}. Below the solidification point, an eutectic exhibits a symmetric phase diagram with two MPBs at compositions $x_0$ and $1-x_0$, separating the two solid phases of the pure materials from the intermediate region of coexistence of both phases. As depicted in the revised phase diagram of Fig. \ref{Diagram}, we thus suggest that the intermediate phase is the Tetra-III=Tetra-I + Tetra-II, with MPBs at approx. 27\% and 73\% FA content, obtained from the onset temperatures of the gap bowing. For completeness and according to recent experimental and theoretical work in FA$_x$MA$_{1-x}$PbBr$_3$ \cite{simen21a} and FAPbI$_3$ \cite{dutta25a}, respectively, we also speculate that the low-T phase of FA-rich compounds may be the Ortho phase with Pnma symmetry of MA-rich compounds. We note that the transition from Tetra-II to (now) Ortho phase for $x\geq0.7$ is characterized by a discontinuity in the gap energy as for MA-rich compounds. In contrast, the Tetra-III/Ortho phase boundary is marked with dashed line in Fig. \ref{Diagram} because its actual position is just conjectural. Regrettably, Raman data were inconclusive in this respect due to a strong PL signal, arising by excitation with the 785-nm laser (1.58 eV) that was above the gap for $x$>0.4 (see S.I. of Ref. [\cite{franc20a}]). 

A highly relevant result from MD simulations performed for a very large supercell is the formation of twin domains with alternating preferential tilt axis parallel to the long tetragonal axis, with 90 deg. domain walls along the $\langle011\rangle$ direction \cite{haine25a}. The aforementioned domains form as a mechanism for strain relaxation and are actually phenomenologically the same as the ferroelastic twin domains appearing in the low-temperature orthorhombic phase upon transition from the tetragonal phase (see, for example, videos of this transformation in MAPbBr$_3$ single crystals in S.I. of Ref. [\cite{ambro22a}]). The presence of ferroelastic domains seems to be key for the activation of the anomalous EP term associated with the FA-rattler modes. This is so because the additional Einstein oscillator of the FA rattlers and the large gap bowing are signatures solely of the Tetra-III and Ortho phases, whereas the homogeneous Cubic and Tetra-II,III phases exhibit a linear gap temperature dependence characteristic of a normal EP coupling mechanism \cite{fasah25a,zacha26b}. For an in-depth discussion of the tight correlation between ferroelastic-domain formation and the immediate appearance of a negative-amplitude Einstein oscillator in the EP term, we refer to Note $\sharp$6 of the S.I. Furthermore, from DFT calculations performed for representative snapshots of MD simulations at 330 K the standard deviation of the valence and conduction band edges was computed as a function of FA composition \cite{haine25a}. The magnitude of the fluctuations is a measure of the strength of the electron-phonon interaction, as experimentally demonstrated by gap fluctuations upon THz or infrared phonon excitation in MAPbI$_3$ \cite{kimxx17a,guoxx20a} and BA$_2$PbI$_4$ \cite{guoxx20a}. The fluctuations are maximal around the MPB, indicating an enhancement of the EP coupling, compatible with the activation of the anomalous EP term, in the phases with twin domains. 
             
%***********************************************

%\section{Conclusions}

In conclusion, by incorporating recent experimental \cite{simen21a} and theoretical \cite{haine25a} results, we have revisited the phase diagram of the FA$_x$MA$_{1-x}$PbI$_3$ system, linking structure, lattice and cation dynamics as well as the electronic band structure, with special emphasis on the understanding of the dependence on temperature of the band gap. In particular, we have elucidated the reason for the pronounced bowing in the gap temperature dependence observed in the low-temperature phases (below ca. 250 K) of the solid solutions for intermediate FA compositions $0.2\leq x \leq0.9$. For this purpose we carried out measurements of the gap pressure coefficient at cryogenic temperatures, using He as pressure transmitting medium, to univocally disentangle the contributions stemming from thermal expansion and electron-phonon interaction. In the Tetra-III an Ortho phases, where the bowing is observed, despite the contribution of the strongly temperature-dependent TE term, an accurate description of the gap bowing is only obtained by invoking the activation of an anomalous EP term represented by an Einstein oscillator with average frequency and coupling strength of $16\pm2$ meV and $-105\pm25$ meV, respectively. In analogy to Cs rattlers \cite{perez23a,fasah25b}, the modes leading to the anomalous coupling are identified as FA-rattlers, that is a combination of inorganic-cage phonons involving octahedral tilting in synchrony with low-frequency FA librations \cite{konto20a,zacha26b}. Although the microscopic mechanism is unknown, the formation of ferroelastic stripe domains with 90-deg. alternating orientation of the preferential tilt axis (coinciding with the long tetragonal or orthorhombic crystal axis) appears to be crucial for the activation of the FA rattlers. These domains naturally form in mixed-cation solid solutions after the transition into the Tetra-III or Ortho phases as a means to minimize the macroscopic strain in the sample. Given the relevance of the gap temperature dependence for the optoelectronic properties of semiconductors in general, and perovskites in particular, its correct assessment is key for the advancement of emergent photovoltaic, efficient light emission, and/or sensing devices, for instance.

%***********************************************************************************************

\section{Supporting Information} Contains details about the synthesis of the mixed-cation perovskites and the PL measurements as a function of temperature, pressure of both simultaneously. It also contains the first-derivative plots, as obtained by point-by-point derivation of the gap-vs-temperature curves, used to disentangle the TE and EP terms from the least-squares fits to the first derivative data points. A detailed analysis of the literature data and the measured gap pressure coefficients so as to determine the TE term as well as the complete set of the least-squares fits to determine the EP terms are also included. Finally, a discussion of the correlation between ferroelastic-domain formation and anomalous EP coupling is provided.

\section{Acknowledgements}

The Spanish "Ministerio de Ciencia, Innovaci\'{o}n y Universidades" (MICIU) through the Agencia Estatal de Investigaci\'{o}n (AEI) and "European Union NextGenerationEU/PRTR" are gratefully acknowledged for its support through grant CEX2023-001263-S (MATRANS42) in the framework of the Spanish Severo Ochoa Centre of Excellence program and the AEI/FEDER(UE) grants PID2021-128924OB-I00 (ISOSCELLES) and PID2024-163010OB-I00 (PV-MENU). The authors also thank the Catalan agency AGAUR for grant 2021-SGR-00444 and the National Network "Red Perovskitas" (MICIU funded). A.F.L. acknowledges a FPI grant BES-2016-076913 from MICIU, co-financed by the European Social Fund, and the PhD programme in Materials Science from Universitat Aut\`{o}noma de Barcelona in which he was enrolled. K.X. acknowledges a fellowship (CSC201806950006) from China Scholarship Council and the PhD programme in Materials Science from Universitat Aut\`{o}noma de Barcelona in which he was enrolled. B.C. thanks the EPSRC for PhD studentship funding via the University of Bath, CSCT CDT (EP/G03768X/1).

\section{Data Availability}
All data generated or analyzed during this study are either included in this published article and its supplementary information files or are available from the corresponding author on reasonable request.

\section{Additional Information}
The authors declare no competing interests.

%\newpage
%\section{References}



\section*{Supplementary material}

# Supporting Information

# Origin of the Temperature-Induced Gap Bowing of Formamidinium-Methylammonium Lead Iodide Perovskites: Role of Cationic Rattlers

Kai Xu,[†] Adrián Francisco-López,[†] Bethan L. Charles,[‡,¶] M. Isabel Alonso,[†] Miquel Garriga,[†] Mark T. Weller,[‡,§] and Alejandro R. Goñi[*,†,‖]

†*Institut de Ciència de Materials de Barcelona, ICMAB-CSIC, Campus UAB, 08193 Bellaterra, Spain*

‡*Dept. of Chemistry & Centre for Sustainable Chemical Technologies, University of Bath, Claverton Down, Bath BA2 7AY, UK*

¶*Dept. of Mechanical Engineering, Queens Building, University of Bristol, Bristol BS8 1TR, UK*

§*Dept. of Chemistry, Cardiff University, Wales CF10 3AT, UK*

‖*ICREA, Passeig Lluís Companys 23, 08010 Barcelona, Spain*

E-mail: goni@icmab.es

## Note ♯0: $FA_xMA_{1-x}PbI_3$ crystal growth

The inverse solubility method of Saidaminov *et al.*[1] was developed to produce crystals of $FA_xMA_{1-x}PbI_3$ across the full composition range. Stoichiometric quantities of MAI, FAI (Great-Cell Solar) and $PbI_2$ (Merck, 99%) were dissolved at 60 °C in dry $\gamma$-Butyrolacetone (Alfa Aesar).

Concentrations of the precursor solutions were kept at 1 M, with an exception for the FA rich materials $FA_xMA_{1-x}PbI_3$ ($x \geq 0.8$), where concentrations were lowered to 0.9 M to allow the FAI to fully dissolve. When the precursor salts and lead halide powders had completely dissolved, the solutions were quickly filtered using Millex GS filters of pore size 0.22 $\mu$m and transferred to clean cylindrical 4 ml glass vials. The solutions were then heated to 110 °C at a rate of 10 °C hour$^{-1}$ using an oil bath, the temperature of which was carefully controlled and monitored using a Stuart SCT1 controller. Solutions were maintained at 110 °C for 2 hours allowing crystallization, the temperature was then raised to 120 °C for a final hour before the remaining solution was filtered and large single crystals were oven dried at 100 °C overnight.

## Note ♯1: Temperature and pressure-dependent photoluminescence (PL) experiments

The PL spectra were excited with the 633 nm line of a He-Ne laser using a very low incident light power below 2 $\mu$W (power density <15 W/cm$^2$) to avoid any photo-degradation of the samples.[2] Spectra were collected using a 20× long working distance objective with NA=0.35 (laser spot of ca. 4 $\mu$m in diameter) and dispersed with a high-resolution LabRam HR800 grating spectrometer equipped with a charge-coupled device detector. PL spectra were corrected for the spectral response of the spectrometer by normalizing each spectrum using the detector and the 600-grooves/mm grating characteristics. Temperature-dependent PL measurements on large single crystals exhibiting flat surfaces were carried out by decreasing the temperature between ca. 365 and 6 K in steps of 5 K, using liquid helium in a gas flow cryostat from CryoVac with optical access that fits under the microscope of the Raman setup.

**Room temperature** high-pressure PL measurements in the range up to 1 GPa were performed employing a gasketed diamond anvil cell (DAC). Anhydrous propanol was used as pressure transmitting medium which ensures perfect hydrostatic conditions[3] up to 4.2 GPa and proved chemically inert to (FA,MA)$PbI_3$. For loading the DAC, small chips with a thickness below ca. 30

$\mu$m were produced by crushing a big single crystal between two glass slides. By close inspection of the debris we were able to pick up small enough, good-quality single crystals, recognized by their flat and shiny surface under the microscope. Small pieces of less than $100\times100$ $\mu$m$^2$ in size were placed into the DAC together with a ruby sphere for pressure calibration.[4] Here we point out that the Inconel gasket was intentionally pre-indented to a fairly large thickness of 120 $\mu$m, before drilling a hole of ca. 250 $\mu$m with a spark-gap machine from EasyLab. The reason was to be able to adjust the pressure with the DAC in steps less than 0.05 GPa, mainly at very low pressures (below 1 GPa). For this purpose an electric motor drive was used to change the pressure in a continuous manner and at low speed (by ca. 0.05 GPa/min).

In contrast, for the high-pressure PL measurements at **low temperatures** down to 4 K a specially designed He-bath cryostat (Konti-IT from Cryovac) was used. The cryostat can allocate the DAC of Syassen/Holzapfel type in its cold bore while granting optical access.[5] Two steel rods designed to fit the threaded rods of the DAC inside the cryostat allow the pressure tuning without removing the DAC from the cryostat. When using the DAC at low temperatures with He as a pressure medium, the loading of the cell proceeds inside the cryostat. The cell is immersed in fluid He, which is pumped to lower its equilibrium temperature till the transition to the superfluid state at 2.17 K occurs. Under this condition, He easily enters into the cell, filling it completely and the DAC is remotely closed. As He instantly crystallizes when subjected to pressure at 2 K, the cell has to be heated well above the melting point to relax non-hydrostatic strain generated by closing the DAC. When cooling down again into the solid phase it is assumed that the pressure on the sample is mostly hydrostatic. To avoid the generation of unwanted uniaxial stress components on the sample, pressure is always changed, if possible, after heating the DAC to a temperature above the solid-liquid phase transition of He. Pressure is monitored in situ using the aforementioned ruby fluorescence method[4] with pertinent temperature correction.[6]

# Note ♯2: Analysis of PL spectra

Figures S1(a-k) of the Supporting information of the work of Francisco-López et al.[7] show the complete set of PL spectra measured at different temperatures between ca. 4 and 385 K for all eleven compositions of $FA_xMA_{1-x}PbI_3$ mixed crystals, spanning the whole compositional range from $x = 0$ to 1 in steps of 10% FA content. In the main manuscript we only show the corresponding spectra cascades for $x$=0.2, 0.5 & 1. The main emission peak observed for all compositions at higher temperature was assigned to free-exciton (FE) recombination.[9–12] Thus, for the quantitative analysis of the PL spectra we performed a lineshape fitting[13] using a Gaussian-Lorentzian cross-product function (lineshape fitting examples were published elsewhere[8]). Its expression reads as:

$$G \times L(\hbar\omega) = A \cdot \frac{\Gamma^2}{4 \cdot s \cdot (E_0 - \hbar\omega)^2 + \Gamma^2} \cdot \exp\left(-4 \cdot \ln 2 \frac{(1-s) \cdot (E_0 - \hbar\omega)^2}{\Gamma^2}\right) \tag{1}$$

where $A$ is the amplitude prefactor, $E_0$ is the peak energy position, $\Gamma$ is the full width at half maximum (FWHM) and $s$ is a weight parameter which takes the value $s = 0$ for pure Gaussian lineshape and $s = 1$ for pure Lorentzian. The values of these four adjustable parameters were obtained as a function of temperature from lineshape fits to the measured PL spectra using Eq. (1). For the cases where there is a coexistence of two phases we used two cross-product functions with independent variable parameter sets. The much more complicated analysis of the wealth of peaks, apparent at low temperatures (below ca. 70 K) for high MA content and ascribed to radiative recombination of bound (acceptor/donor) exciton complexes, was published elsewhere.[14] For all compositions the peak energy position $E_0$ which can be taken as representative of the band gap of the material, exhibits a clear and pronounced bowing at intermediate to low temperatures. In order to understand the origin of this bowing, in the following we will disentangle the contributions to the gap temperature dependence stemming from thermal expansion and electron-phonon interaction at selected temperatures and for some of the eleven compositions.

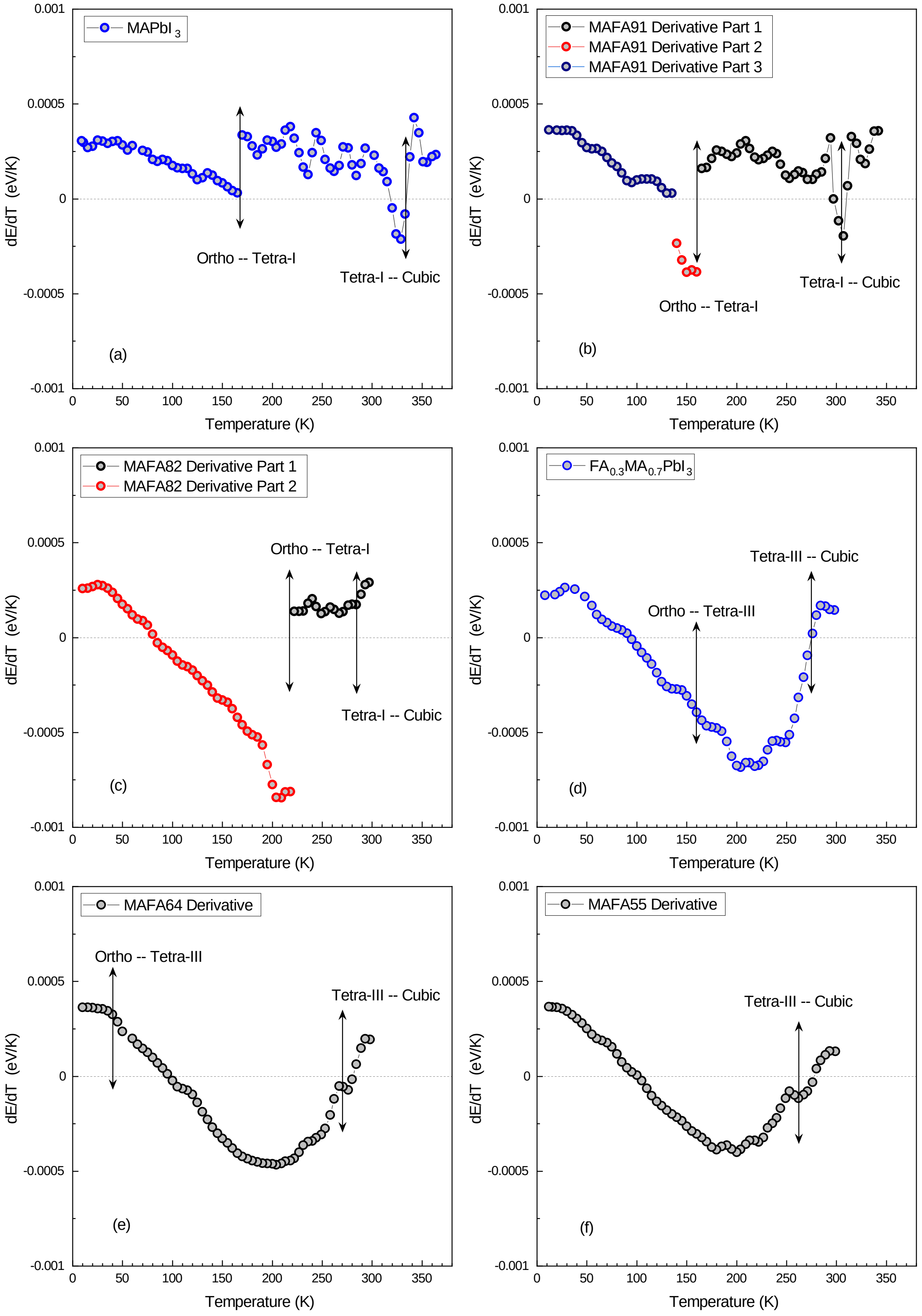
MAPbI 3
Ortho -- Tetra-I
Tetra-I -- Cubic
(a)
MAFA91 Derivative Part 1
MAFA91 Derivative Part 2
MAFA91 Derivative Part 3
Tetra-I -- Cubic
Ortho -- Tetra-I
(b)
MAFA82 Derivative Part 1
MAFA82 Derivative Part 2
Ortho -- Tetra-I
Tetra-I -- Cubic
(c)
$FA_{0.3}MA_{0.7}PbI_3$
Tetra-III -- Cubic
Ortho -- Tetra-III
(d)
MAFA64 Derivative
Ortho -- Tetra-III
Tetra-III -- Cubic
(e)
MAFA55 Derivative
Tetra-III -- Cubic
(f)
dE/dT (eV/K)
Temperature (K)
0.001
0.0005
0
-0.0005
-0.001
0
50
100
150
200
250
300
350

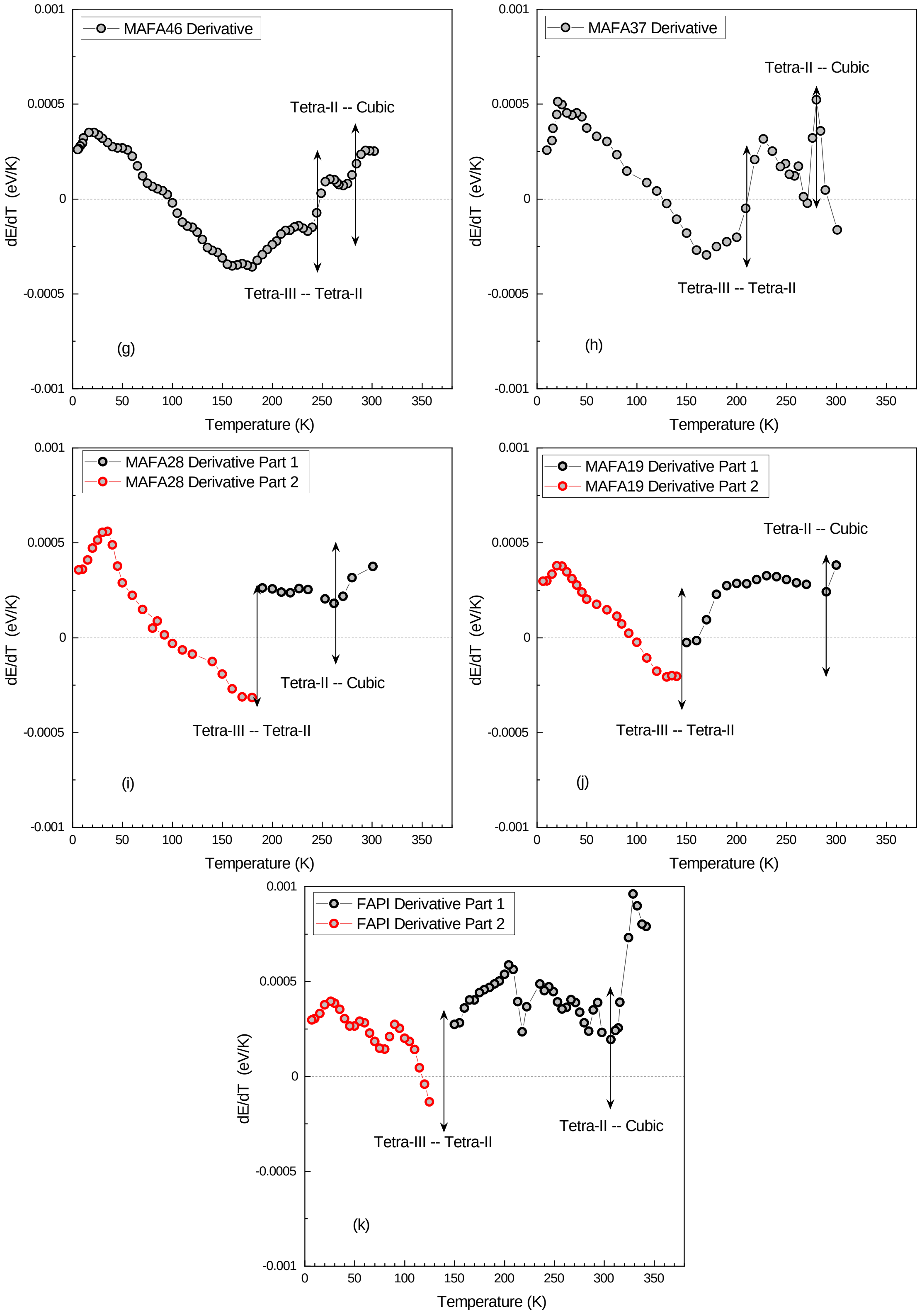


Figure S 0: First derivative of the PL peak energy $E_0$ with respect to temperature, plotted as a function of temperature, for the $FA_xMA_{1-x}PbI_3$ series with (a) $x = 0$, (b) 0.1, (c) 0.2, (d) 0.3, (e) 0.4, (f) 0.5, (g) 0.6, (h) 0.7, (i) 0.8, (j) 0.9, and (k) $x = 1$. The temperatures at which different structural phase transitions occur are indicated by arrows.

# Note ♯3: Temperature dependence of the gap energy & its first derivative

The values of the fitting parameter corresponding to the peak energy $E_0$, which we take for the gap energy, as obtained from the line shape fits using Eq. (1), is plotted as a function of temperature in Fig. 2 of the main manuscript for the eleven compositions of $FA_xMA_{1-x}PbI_3$ studied. In Figs. S0(a)-(k) we illustrate the procedure to construct the graphs of the derivative of $E_0$ over temperature plotted as a function of temperature for each FA concentration $x \in [0,1]$ in steps of 0.1, respectively. In order to avoid the amplification of the scatter of the data points when taking the first derivative numerically and point-by-point, we have previously smoothed the E-vs-T curves (see procedure, for example, in the Supporting Information of Ref. [15]. These first-derivative plots will be very important for disentangling the effects of thermal expansion and electron-phonon coupling, as well as for the discussion of the structural behavior of the $FA_xMA_{1-x}PbI_3$ solid-solution series. The temperatures at which different structural phase transitions occur are indicated by arrows. For low FA concentrations the temperature of the phase transition between the cubic and the tetra-I phase (the same procedure is valid for the cubic-tetra-II transition for high FA contents), which in the $E_0$-vs.-$T$ curve corresponds to a zigzag type behavior, is extracted from the zero of the first derivative curve, as indicated in Fig. S0(a), for example. In the same figure, we also show the determination of the transition temperature for the tetra-I to orthorhombic transformation, which simply corresponds to the temperature at which the jump in the first derivative occurs. For the smooth transition between the cubic and the tetra-III phase, taking place for intermediate FA concentrations, the transition temperature is obtained again from the crossing point of the first derivative curve and the temperature axis (see example in Fig. S0(c)). This corresponds to the inflection point in the $E_0$-vs.-$T$ curve, which is less straightforward to be determined.

# Note ♯4: The thermal expansion (TE) term

The derivative of the gap over temperature contains two terms; one accounts for thermal expansion effects (TE) and the other corresponds to the renormalization directly caused by electron-phonon interaction (EP), which includes the Debye-Waller and self-energy corrections:[16–18]

$$\frac{dE_g}{dT} = \left[\frac{\partial E_g}{\partial T}\right]_{TE} + \left[\frac{\partial E_g}{\partial T}\right]_{EP}. \tag{2}$$

The effect on the gap due to the contraction of the lattice with decreasing temperature is intimately related to the response of the electronic band structure upon application of external hydrostatic pressure. It thus holds[16,17,19]

$$\left[\frac{\partial E_g}{\partial T}\right]_{TE} = -\alpha_V \cdot B_0 \cdot \frac{dE_g}{dP}, \tag{3}$$

where $\alpha_V$ is the volumetric expansion coefficient, $B_0$ is the bulk modulus, i.e. the inverse of the compressibility, and $\frac{dE_g}{dP}$ is the pressure coefficient of the gap, which can be determined from high pressure experiments. In the entire temperature range of the experiments, $\alpha_V$ is positive and the sign of Eq. (3) for the thermal expansion contribution is determined by the sign of the pressure coefficient. For most metal halide perovskites (MHPs) direct gaps $\frac{dE_g}{dP}$ is negative around room temperature, hence, thermal expansion causes a gap increase.[20] The negative sign of $\frac{dE_g}{dP}$ arises from the inverted atomic orbital character of the states at the top and bottom of the valence and conduction band of lead halide perovskites, respectively, as explained elsewhere.[8] At lower temperatures than ambient the situation can fundamentally change, where positive gap pressure coefficients are measured, which lead to a reduction of the gap with temperature. In fact, this is the case of the Tetra-III phase of the $FA_xMA_{1-x}PbI_3$ mixed crystals at intermediate temperatures for which the pronounced gap bowing is observed.

For the determination of the TE term, according to Eq. (3), we first need to search the literature looking for values of the volumetric thermal expansion coefficient of the solid sotutions as a function of temperature. Fortunately, we found data for the two extreme compositions ($MAPbI_3$

& $FAPbI_3$) and the intermediate one ($FA_{0.5}MA_{0.5}PbI_3$). For the rest of the compositions we were forced to interpolate the $\alpha_V$ data making sensible assumptions, as indicated below. In Table S0 we summarize the thermal expansion coefficient values $\alpha_V(T)$ from the cited sources, indicating the temperature range for which each value is valid. We point out that the listed $\alpha_V(T)$ were obtained from the measured unit-cell volumes $V(T)$ by computing the slope $\frac{1}{V(T)} \cdot \frac{V(T+\Delta T)-V(T)}{\Delta T}$ stepwise for each temperature range indicated in Table S0.

Table S 0: Computed volumetric thermal expansion coefficient $\alpha_V(T)$ as a function of temperature for $MAPbI_3$, $FA_{0.5}MA_{0.5}PbI_3$ and $FAPbI_3$ obtained from the measured volume-vs-temperature data of the cited literature. Also indicated are the different temperature ranges of validity of $\alpha_V$ and the corresponding stable phase.

| $x$ | $\alpha_V$ ($10^{-4}$ K$^{-1}$) | Phase | Temperature range (K) | Reference |
|---|---|---|---|---|
| 0 | 1.41 | Cubic | 325 - 375 | [21] |
| | 1.27 | Tetra-I | 160 - 320 | [22] |
| | 1.56 | Ortho | 140 - 155 | [23] |
| | 1.14 | Ortho | 85 - 140 | [23] |
| | 1.01 | Ortho | 55 - 85 | [23] |
| | 0.81 | Ortho | 30 - 55 | [23] |
| | 0.26 | Ortho | 10 - 30 | [23] |
| 0.5 | 1.54 | Cubic | 275 - 300 | [24] |
| | 2.30 | Tetra-III | 150 - 275 | [24] |
| 1 | 1.08 | Cubic | 300 - 500 | [25] |
| | 1.90 | Tetra-II | 150 - 300 | [25] |
| | 1.75 | Tetra-III | 90 - 150 | [25] |

We now turn to the discussion of the bulk modulus $B_0$ and its possible dependence on composition and temperature. With respect to the temperature dependence of $B_0$, the situation is very unsatisfactory, since, to the best of our knowledge, there are almost no data available, not even for the archetypal perovskite $MAPbI_3$. The only solution is to use, for each of the phases the material adopts as a function of temperature, the $B_0$ value obtained for the corresponding high-pressure phase at room temperature. This assumption is supported by temperature-dependent de-

terminations of the Young modulus of $FAPbI_3$,[26] which indicate that the elastic properties of that perovskite are essentially constant in the entire temperature range of stability of each crystalline phase. Furthermore, a study of the elastic properties of four MA metal halides, namely $MAPbI_3$, $MAPbBr_3$ and $MAPb(Br_xI_{1-x})_3$ with $x$=0.2 & 0.6 has shown that there is no clear systematics in composition for the bulk moduli of these compounds, being all within experimental uncertainty.[27] In conclusion, to compute the TE term we have used the same bulk modulus value for all cubic, all tetragonal and all orthorhombic phases, that is $B_0$ values independent of temperature and composition. The actual numbers correspond to the accurate values determined for $MAPbI_3$ at high pressures but room temperature:[28] 16(3) GPa for the tetragonal-I phase stable at ambient, 13(5) GPa for the cubic phase stable between 0.4 and 3 GPa and 30(2) GPa for the orthorhombic phase stable above 3 GPa. Numbers in parentheses are the error bars.

Finally, according to Eq. (3), the gap pressure coefficients at given temperatures, $\frac{dE_g}{dP}(T)$, are needed for the assessment of the TE term at that temperatures. For this purpose, we run high-pressure PL experiments at cryogenic temperatures using He as pressure-transmitting medium. Since these are extremely complex and time-consuming experiments, we performed the measurements for only seven of the eleven studied solid-solution compositions and for a small set of representative temperatures. The gap pressure coefficient of Eq. (3) is determined from the slope of the energy versus pressure curves, where the data points were previously obtained from fits to the PL spectra recorded at different pressures.

Figure S1 shows the variation with pressure of the gap energy of $MAPbI_3$, as obtained from PL line-shape fits of the PL spectra measured at low pressure and at the selected temperatures. The slopes of the linear fits to the data points (red solid lines in Fig. S1) are listed in Table S1 together with the values of $B_0$ and $\alpha_V$ employed to calculate the thermal expansion term TE in the case of $MAPbI_3$. We have also included the only available data, other than the ones reported here, concerning the pressure coefficient of the gap in the orthorhombic phase.[29]

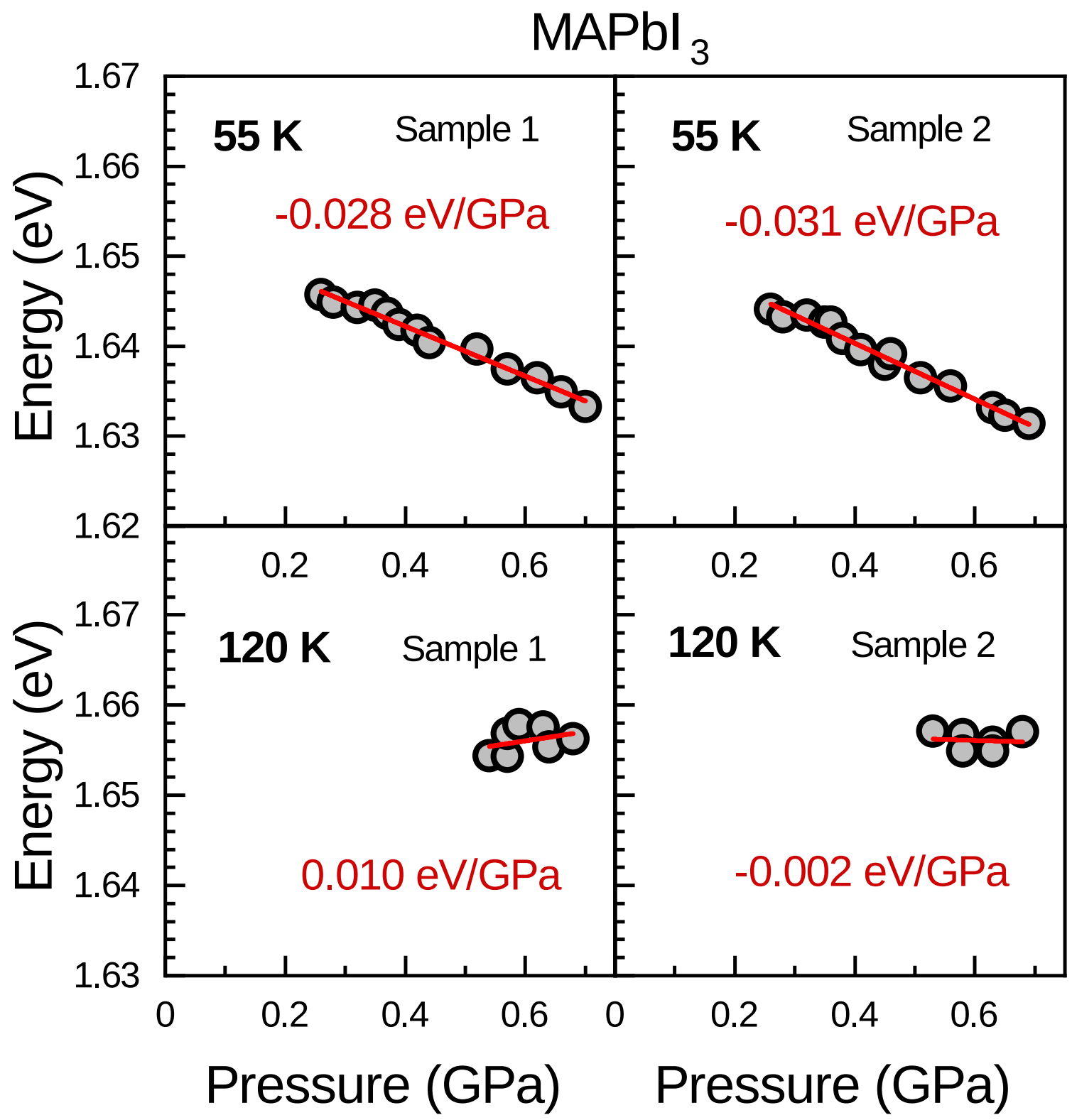


Figure S 1: The energy of the gap extracted from the PL peak maximum (symbols) plotted as a function of pressure and for different temperatures, as indicated, in the case of $MAPbI_3$. The red lines represent a fit to the data points using a linear function. The slope values are also indicated.

Table S 1: Computed thermal expansion term $\left[\frac{\partial E_g}{\partial T}\right]_{TE}$ from the values of the volumetric thermal expansion coefficient $\alpha_V(T)$, bulk modulus $B_0$ and gap pressure coefficient $\frac{dE_g}{dP}$ for $MAPbI_3$ and different temperatures $T$. Numbers in parentheses are the error bars. The asterisks indicate that the data is from Ref. [29] and the dagger indicates that $B_0$ is from the room-temperature high-pressure phase (Ref. [28]).

| | **$MAPbI_3$** | | | | |
|---|---|---|---|---|---|
| Phase | $T$ (K) | $\alpha_V$ ($10^{-4}$ $K^{-1}$) | $B_0$ (GPa) | $\frac{dE_g}{dP}$ (eV/GPa) | $\left[\frac{\partial E_g}{\partial T}\right]_{TE}$ ($10^{-4}$ eV/K) |
| Ortho | 40* | 0.81(5) | 30(6) | -0.036(6)* | 0.87(12) |
| | 55 | 1.01(5) | 30(6) | -0.028(3) | 0.85(9) |
| | 55 | 1.01(5) | 30(6) | -0.031(3) | 0.94(9) |
| | 80* | 1.01(5) | 30(6) | -0.029(1)* | 0.88(6) |
| | 120 | 1.14(5) | 30(6) | 0.010(10) | -0.34(15) |
| | 120 | 1.14(5) | 30(6) | -0.002(10) | 0.07(15) |
| | 120* | 1.14(5) | 30(6) | -0.013(2) | 0.44(5) |
| Tetra-I | 300 | 1.27(5) | 16(3) | -0.050(10) | 1.02(4) |
| Cubic | 350 | 1.41(5) | 13(5)† | -0.012(10) | 0.22(6) |

Figure S2 shows the variation with pressure of the gap energy of $FA_{0.2}MA_{0.8}PbI_3$, as obtained from PL line-shape fits of the PL spectra measured at low pressure and at the selected temperatures. The slopes of the linear fits to the data points (red solid lines in Fig. S2) are listed in Table S2 together with the values of $B_0$ and $\alpha_V$ employed to calculate the thermal expansion term TE.

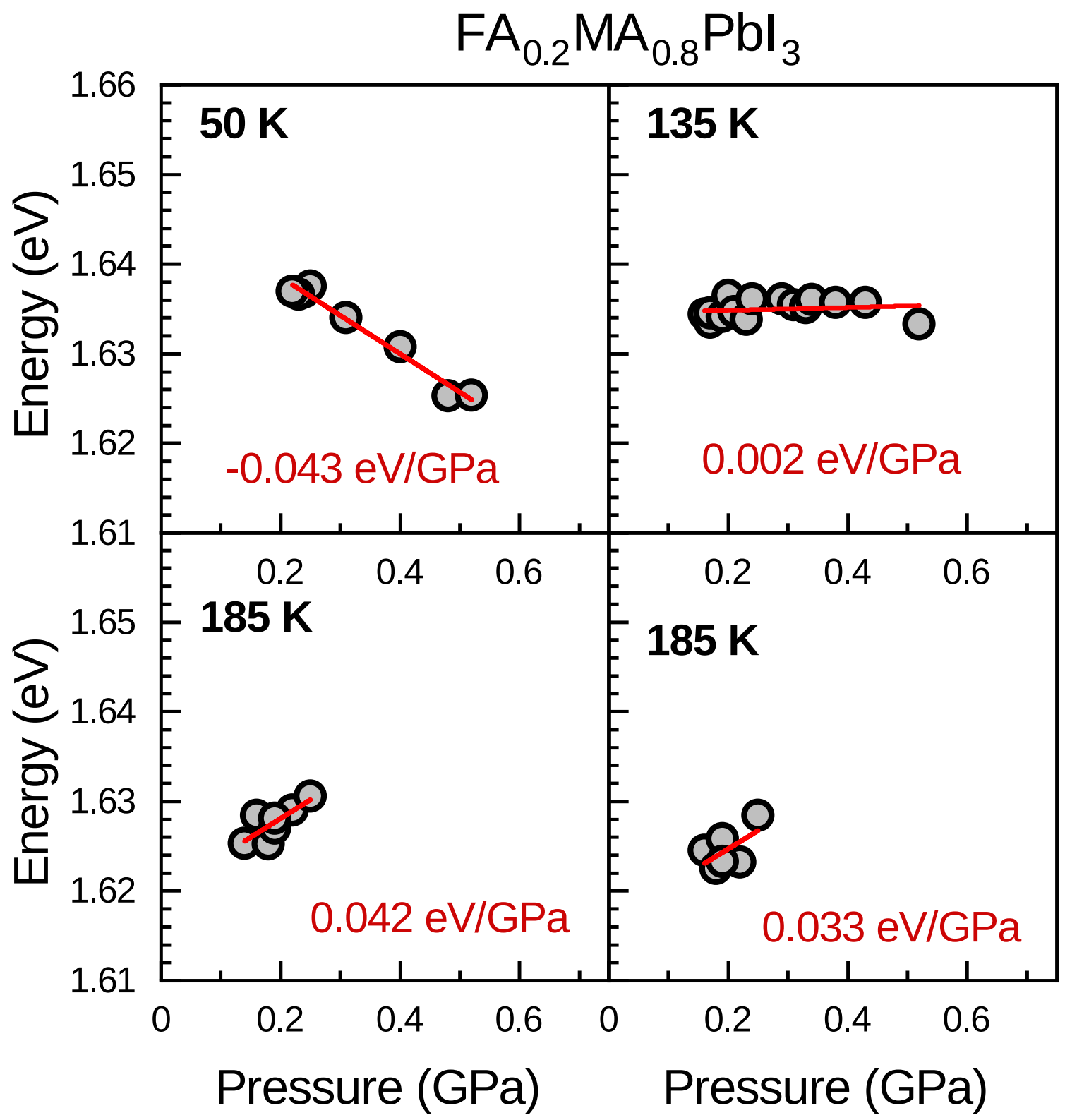


Figure S 2: The energy of the gap extracted from the PL peak maximum (symbols) plotted as a function of pressure and for different temperatures, as indicated, in the case of $FA_{0.2}MA_{0.8}PbI_3$. The red lines represent a fit to the data points using a linear function. The slope values are also indicated.

Table S 2: Computed thermal expansion term $\left[\frac{\partial E_g}{\partial T}\right]_{TE}$ from the values of the volumetric thermal expansion coefficient $\alpha_V(T)$, bulk modulus $B_0$ and gap pressure coefficient $\frac{dE_g}{dP}$ for $FA_{0.2}MA_{0.8}PbI_3$ and different temperatures $T$. Numbers in parentheses are the error bars.

| | **$FA_{0.2}MA_{0.8}PbI_3$** | | | | |
|---|---|---|---|---|---|
| Phase | $T$ (K) | $\alpha_V$ ($10^{-4}$ $K^{-1}$) | $B_0$ (GPa) | $\frac{dE_g}{dP}$ (eV/GPa) | $\left[\frac{\partial E_g}{\partial T}\right]_{TE}$ ($10^{-4}$ eV/K) |
| Ortho | 50 | 1.21(5) | 30(6) | -0.043(6) | 1.56(20) |
| | 135 | 1.37(5) | 30(6) | 0.002(6) | -0.07(12) |
| | 180 | 1.87(5) | 30(6) | 0.042(10) | -1.83(30) |
| | 180 | 1.87(5) | 30(6) | 0.033(10) | -2.35(30) |
| Cubic | 300 | 1.69(5) | 13(5) | -0.048(8) | 1.05(20) |
| | 300 | 1.69(5) | 13(5) | -0.044(10) | 0.97(20) |
| | 300 | 1.69(5) | 13(5) | -0.046(3) | 1.01(20) |
| | 300 | 1.69(5) | 13(5) | -0.038(3) | 0.83(15) |

Figure S3 shows the variation with pressure of the gap energy of $FA_{0.4}MA_{0.6}PbI_3$, as obtained from PL line-shape fits of the PL spectra measured at low pressure and at the selected temperatures. The slopes of the linear fits to the data points (red solid lines in Fig. S3) are listed in Table S3 together with the values of $B_0$ and $\alpha_V$ employed to calculate the thermal expansion term TE.

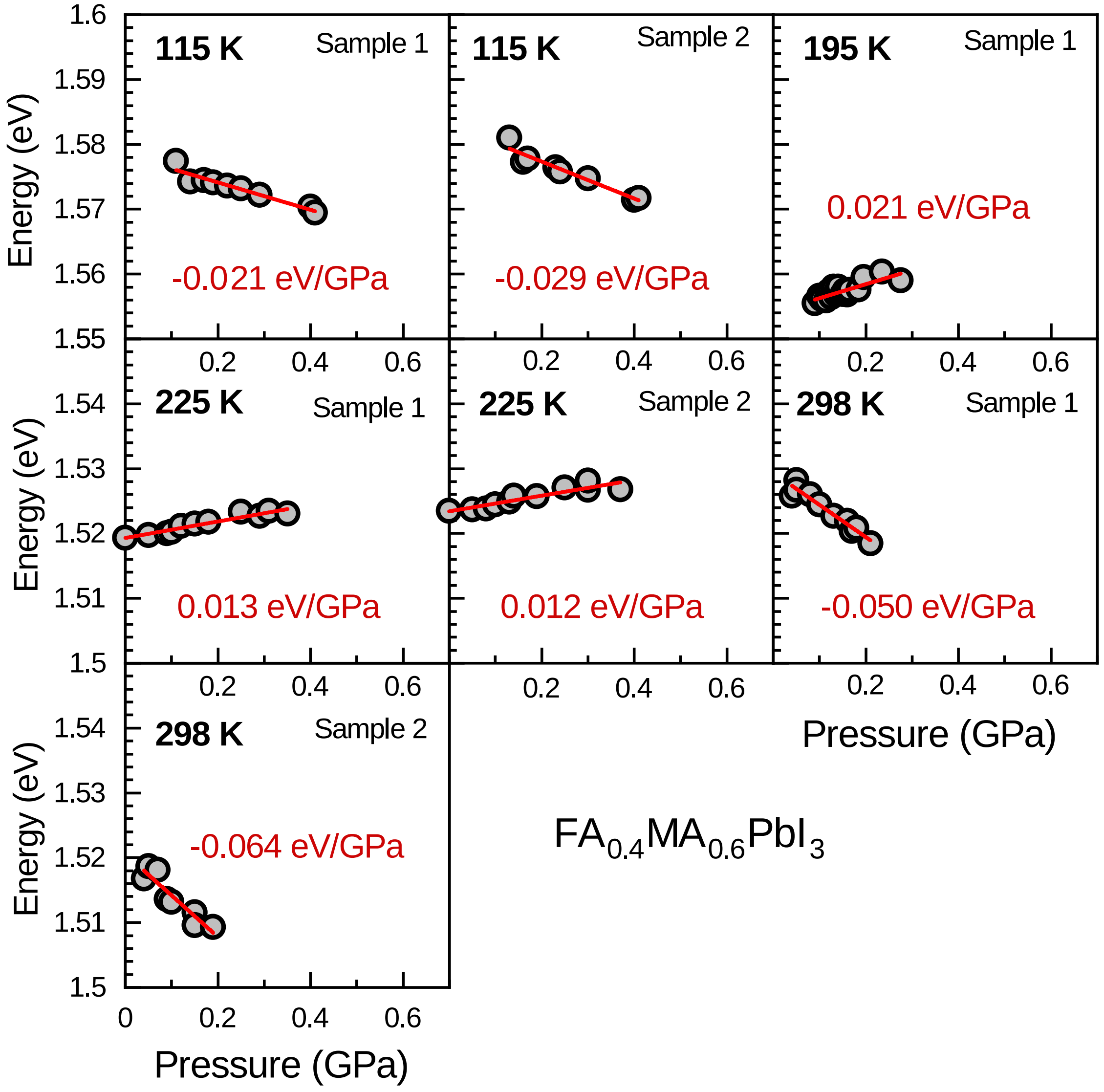


Figure S 3: The energy of the gap extracted from the PL peak maximum (symbols) plotted as a function of pressure and for different temperatures, as indicated, in the case of $FA_{0.4}MA_{0.6}PbI_3$. The red lines represent a fit to the data points using a linear function. The slope values are also indicated.

Table S 3: Computed thermal expansion term $\left[\frac{\partial E_g}{\partial T}\right]_{TE}$ from the values of the volumetric thermal expansion coefficient $\alpha_V(T)$, bulk modulus $B_0$ and gap pressure coefficient $\frac{dE_g}{dP}$ for $FA_{0.4}MA_{0.6}PbI_3$ and different temperatures $T$. Numbers in parentheses are the error bars.

| | **$FA_{0.4}MA_{0.6}PbI_3$** | | | | |
|---|---|---|---|---|---|
| Phase | $T$ (K) | $\alpha_V$ ($10^{-4}$ K$^{-1}$) | $B_0$ (GPa) | $\frac{dE_g}{dP}$ (eV/GPa) | $\left[\frac{\partial E_g}{\partial T}\right]_{TE}$ ($10^{-4}$ eV/K) |
| Tetra-III | 115 | 1.90(5) | 15(6) | -0.021(5) | 0.60(20) |
| | 115 | 1.90(5) | 15(6) | -0.029(5) | 0.83(12) |
| | 195 | 2.30(5) | 15(6) | 0.021(5) | -0.73(20) |
| | 225 | 2.30(5) | 15(6) | 0.013(10) | -0.45(30) |
| | 225 | 2.30(5) | 15(6) | 0.012(10) | -0.42(30) |
| Cubic | 298 | 1.54(5) | 12(5) | -0.050(3) | 0.92(10) |
| | 298 | 1.54(5) | 12(5) | -0.064(3) | 1.18(10) |

Figure S4 shows the variation with pressure of the gap energy of $FA_{0.5}MA_{0.5}PbI_3$, as obtained from PL line-shape fits of the PL spectra measured at low pressure and at the selected temperatures. The slopes of the linear fits to the data points (red solid lines in Fig. S4) are listed in Table S4 together with the values of $B_0$ and $\alpha_V$ employed to calculate the thermal expansion term TE.

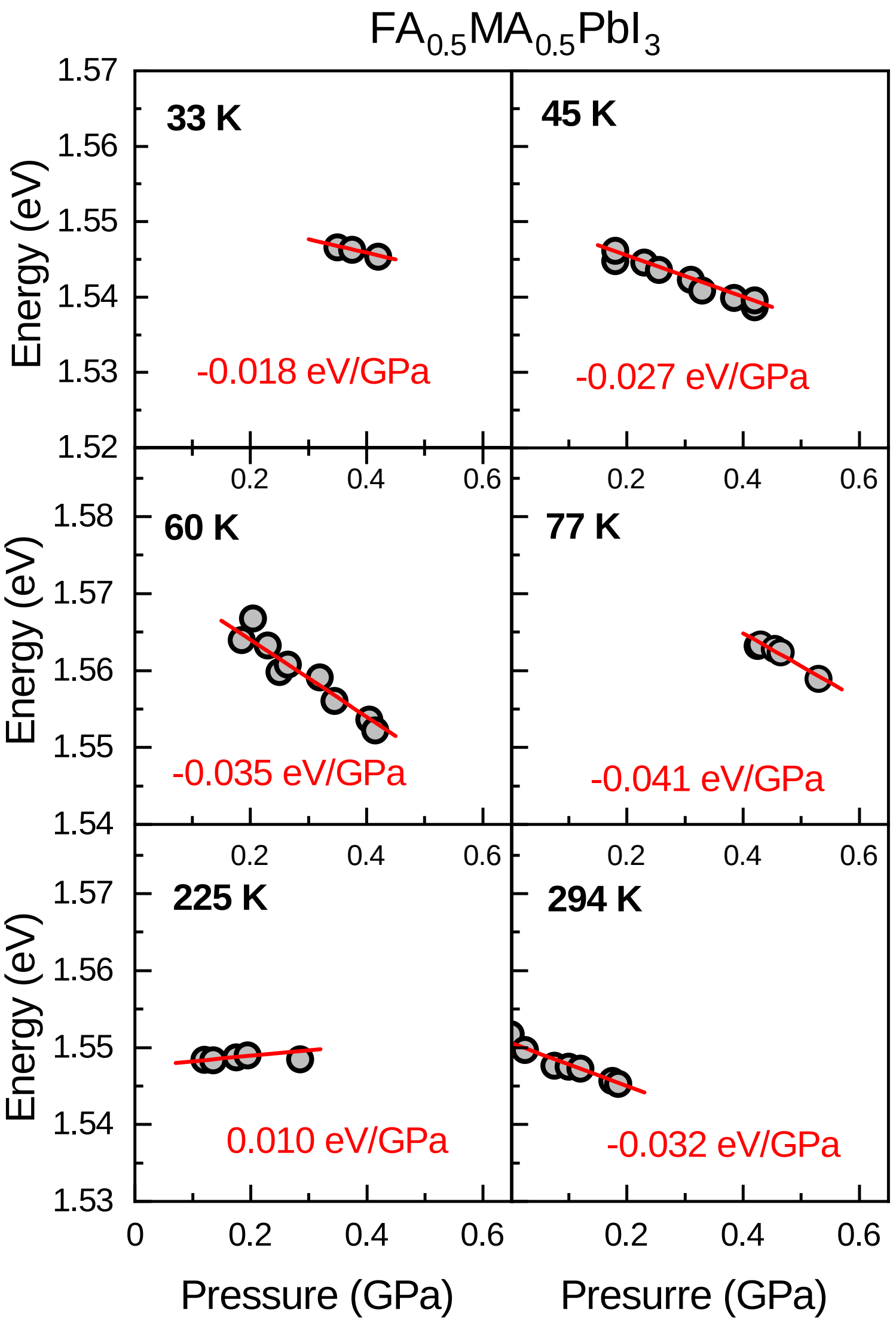


Figure S 4: The energy of the gap extracted from the PL peak maximum (symbols) plotted as a function of pressure and for different temperatures, as indicated, in the case of $FA_{0.5}MA_{0.5}PbI_3$. The red lines represent a fit to the data points using a linear function. The slope values are also indicated.

Table S 4: Computed thermal expansion term $\left[\frac{\partial E_g}{\partial T}\right]_{TE}$ from the values of the volumetric thermal expansion coefficient $\alpha_V(T)$, bulk modulus $B_0$ and gap pressure coefficient $\frac{dE_g}{dP}$ for $FA_{0.5}MA_{0.5}PbI_3$ and different temperatures $T$. Numbers in parentheses are the error bars. The asterisk indicates that the data is for the Tetra-III phase of $FAPbI_3$[25] and the dagger indicates the data is from Ref. [24].

| | **$FA_{0.5}MA_{0.5}PbI_3$** | | | | |
|---|---|---|---|---|---|
| Phase | $T$ (K) | $\alpha_V$ ($10^{-4}$ K$^{-1}$) | $B_0$ (GPa) | $\frac{dE_g}{dP}$ (eV/GPa) | $\left[\frac{\partial E_g}{\partial T}\right]_{TE}$ ($10^{-4}$ eV/K) |
| Tetra-III | 33 | 0.32(8) | 15(7) | -0.018(15) | 0.09(8) |
| | 45 | 0.99(8) | 15(7) | -0.027(15) | 0.4 (3) |
| | 60 | 1.23(8) | 15(7) | -0.035(8) | 0.65(35) |
| | 77 | 1.23(8) | 15(7) | -0.038(8) | 0.7 (4) |
| | 225 | 1.75(8)$^*$ | 15(7) | 0.010(8) | -0.25(25) |
| Cubic | 294 | 1.54(5)$^\dagger$ | 12(5) | -0.032(15) | 0.6(3) |

Figure S5 shows the variation with pressure of the gap energy of $FA_{0.6}MA_{0.4}PbI_3$, as obtained from PL line-shape fits of the PL spectra measured at low pressure and at the selected temperatures. The slopes of the linear fits to the data points (red solid lines in Fig. S5) are listed in Table S5 together with the values of $B_0$ and $\alpha_V$ employed to calculate the thermal expansion term TE.

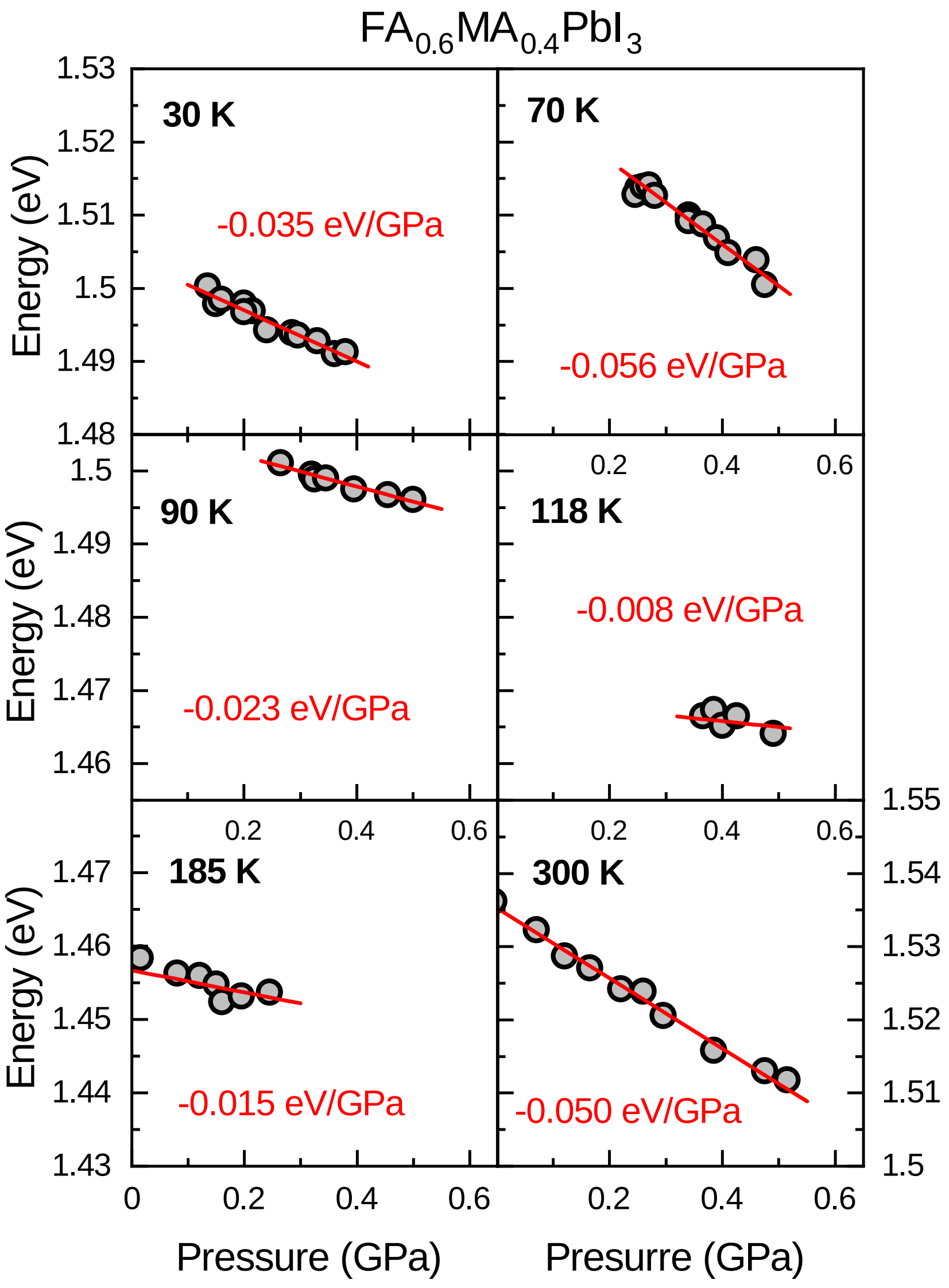


Figure S 5: The energy of the gap extracted from the PL peak maximum (symbols) plotted as a function of pressure and for different temperatures, as indicated, in the case of $FA_{0.6}MA_{0.4}PbI_3$. The red lines represent a fit to the data points using a linear function. The slope values are also indicated.

Table S 5: Computed thermal expansion term $\left[\frac{\partial E_g}{\partial T}\right]_{TE}$ from the values of the volumetric thermal expansion coefficient $\alpha_V(T)$, bulk modulus $B_0$ and gap pressure coefficient $\frac{dE_g}{dP}$ for $FA_{0.6}MA_{0.4}PbI_3$ and different temperatures $T$. Numbers in parentheses are the error bars.

| | **$FA_{0.6}MA_{0.4}PbI_3$** | | | | |
|---|---|---|---|---|---|
| Phase | $T$ (K) | $\alpha_V$ ($10^{-4}$ K$^{-1}$) | $B_0$ (GPa) | $\frac{dE_g}{dP}$ (eV/GPa) | $\left[\frac{\partial E_g}{\partial T}\right]_{TE}$ ($10^{-4}$ eV/K) |
| Tetra-III | 30 | 0.99(8) | 15(7) | -0.035(4) | 0.5(1) |
| | 70 | 1.23(8) | 15(7) | -0.056(10) | 1.0(3) |
| | 90 | 1.75(8) | 15(7) | -0.023(6) | 0.6(2) |
| | 118 | 1.75(8) | 15(7) | -0.008(6) | 0.2(2) |
| Tetra-II | 185 | 1.90(6) | 12(7) | -0.015(5) | 0.35(20) |
| | 207 | 1.90(6) | 12(7) | -0.003(5) | 0.07(6) |
| Cubic | 300 | 1.54(5) | 12(5) | -0.060(10) | 1.1(2) |
| | 300 | 1.54(5) | 12(5) | -0.050(5) | 0.9(1) |
| | 300 | 1.54(5) | 12(5) | -0.054(8) | 1.0(2) |

Figure S6 shows the variation with pressure of the gap energy of $FA_{0.8}MA_{0.2}PbI_3$, as obtained from PL line-shape fits of the PL spectra measured at low pressure and at the selected temperatures. The slopes of the linear fits to the data points (red solid lines in Fig. S6) are listed in Table S6 together with the values of $B_0$ and $\alpha_V$ employed to calculate the thermal expansion term TE.

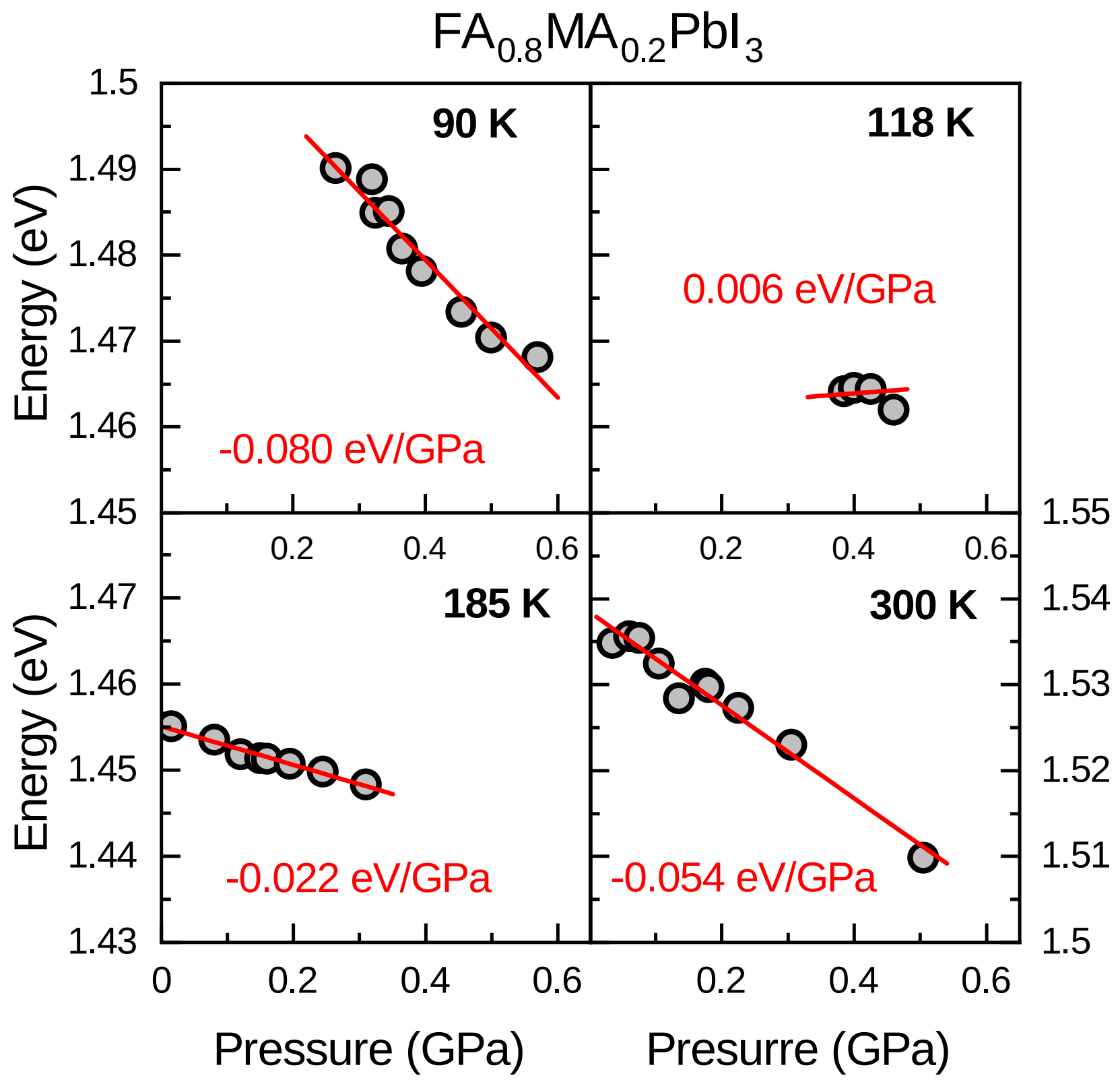


Figure S 6: The energy of the gap extracted from the PL peak maximum (symbols) plotted as a function of pressure and for different temperatures, as indicated, in the case of $FA_{0.8}MA_{0.2}PbI_3$. The red lines represent a fit to the data points using a linear function. The slope values are also indicated.

Table S 6: Computed thermal expansion term $\left[\frac{\partial E_g}{\partial T}\right]_{TE}$ from the values of the volumetric thermal expansion coefficient $\alpha_V(T)$, bulk modulus $B_0$ and gap pressure coefficient $\frac{dE_g}{dP}$ for $FA_{0.8}MA_{0.2}PbI_3$ and different temperatures $T$. Numbers in parentheses are the error bars.

| | $\mathbf{FA_{0.8}MA_{0.2}PbI_3}$ | | | | |
|---|---|---|---|---|---|
| Phase | $T$ (K) | $\alpha_V$ ($10^{-4}$ $K^{-1}$) | $B_0$ (GPa) | $\frac{dE_g}{dP}$ (eV/GPa) | $\left[\frac{\partial E_g}{\partial T}\right]_{TE}$ ($10^{-4}$ eV/K) |
| Tetra-III | 90<br>118 | 1.75(8)<br>1.90(8) | 15(7)<br>12(7) | -0.080(15)<br>0.006(8) | 2.1 (5)<br>-0.14(3) |
| Tetra-II | 185 | 1.90(6) | 12(7) | -0.022(5) | 0.5 (2) |
| Cubic | 300 | 1.08(5) | 12(5) | -0.054(5) | 0.7(1) |

Figure S7 shows the variation with pressure of the gap energy of $FA_{0.9}MA_{0.1}PbI_3$, as obtained

from PL line-shape fits of the PL spectra measured at low pressure and at the selected temperatures. The slopes of the linear fits to the data points (red solid lines in Fig. S7) are listed in Table S7 together with the values of $B_0$ and $\alpha_V$ employed to calculate the thermal expansion term TE.

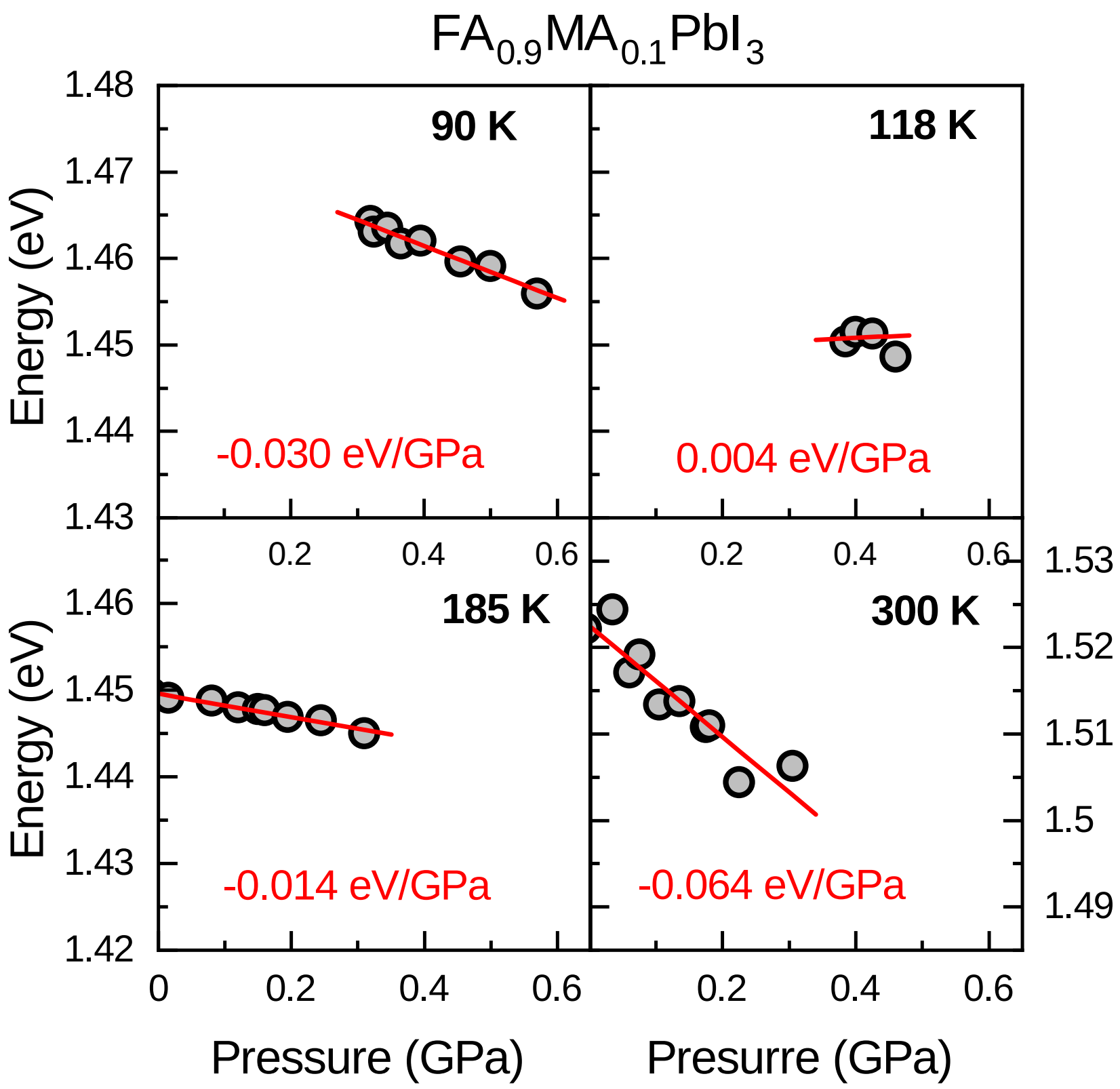


Figure S 7: The energy of the gap extracted from the PL peak maximum (symbols) plotted as a function of pressure and for different temperatures, as indicated, in the case of $FA_{0.9}MA_{0.1}PbI_3$. The red lines represent a fit to the data points using a linear function. The slope values are also indicated.

Table S 7: Computed thermal expansion term $\left[\frac{\partial E_g}{\partial T}\right]_{TE}$ from the values of the volumetric thermal expansion coefficient $\alpha_V(T)$, bulk modulus $B_0$ and gap pressure coefficient $\frac{dE_g}{dP}$ for $FA_{0.9}MA_{0.1}PbI_3$ and different temperatures $T$. Numbers in parentheses are the error bars.

| | **$FA_{0.9}MA_{0.1}PbI_3$** | | | | |
|---|---|---|---|---|---|
| Phase | $T$ (K) | $\alpha_V$ ($10^{-4}$ $K^{-1}$) | $B_0$ (GPa) | $\frac{dE_g}{dP}$ (eV/GPa) | $\left[\frac{\partial E_g}{\partial T}\right]_{TE}$ ($10^{-4}$ eV/K) |
| Tetra-III | 90 | 1.75(8) | 15(7) | -0.030(8) | 0.8 (2) |
| | 118 | 1.90(8) | 12(7) | 0.004(10) | -0.09(2) |
| Tetra-II | 185 | 1.90(6) | 12(7) | -0.014(5) | 0.3 (1) |
| Cubic | 300 | 1.08(5) | 12(5) | -0.064(6) | 0.8(1) |

# Note ♯5: The electron-phonon (EP) coupling term

Gopalan *et al.*[17] derived an expression for the energy shift of the gap induced by temperature, through electron-phonon interaction, of any electronic state $E_{n\mathbf{k}}$ with band index $n$ and wavevector $\mathbf{k}$, for which all phonon modes of the branch $j$, wavevector $\mathbf{q}$ and frequency $\omega_{j\mathbf{q}}$ contribute:

$$\Delta E_{n\mathbf{k}}(T) = \sum_{j\mathbf{q}} \frac{\partial E_{n\mathbf{k}}}{\partial n_{j\mathbf{q}}} \left( n_{j\mathbf{q}}(T) + \frac{1}{2} \right), \tag{4}$$

where $n_{j\mathbf{q}} = \left(e^{\beta\hbar\omega_{j\mathbf{q}}} - 1\right)^{-1}$ is the Bose-Einstein phonon occupation factor with $\beta = \frac{1}{k_B T}$. Importantly, the complex interaction coefficients $\frac{\partial E_{n\mathbf{k}}}{\partial n_{j\mathbf{q}}}$ contribute *both* to the energy shift of the bands as well as to a lifetime broadening of the electronic states (linewidths). Obviously, the sign of the renormalization of a gap is thus determined by the difference in magnitude and sign of the respective energy shift of valence and conduction band.

By invoking energy conservation, the summation in Eq. (4) transforms into an integral over the

phonon frequencies:[17]

$$\Delta E_{n\mathbf{k}}(T) = \int_0^\infty d\omega \cdot g^2F(n,\mathbf{k},\omega) \cdot \left(n_{j\mathbf{q}}(T) + \frac{1}{2}\right)$$
$$g^2F(n,\mathbf{k},\omega) = \sum_{j\mathbf{q}} \frac{\partial E_{n\mathbf{k}}}{\partial n_{j\mathbf{q}}} \delta(\omega - \omega_{j\mathbf{q}}). \tag{5}$$

The function $g^2F(n,\mathbf{k},\omega)$ is the so-called electron-phonon spectral function and is essentially the phonon density of states (DOS) appropriately weighted by electron-phonon matrix elements. As such, the spectral function is temperature independent, which means that the temperature dependence of the electron-phonon contribution to the gap shift arises solely from the Bose-Einstein occupation factor $n_{j\mathbf{q}}(T)$.

Equations. (4) and (5) indicate that the main contributions to the electron-phonon renormalization of the gap arise from peaks in the DOS. In fact, this is at the origin of the Einstein-oscillator model introduced by Cardona and coworkers,[18,30,31] which approximate the $\frac{\partial E_{n\mathbf{k}}}{\partial n_{j\mathbf{q}}}$ coefficients by effective electron-phonon interaction parameters $A_i$ for phonons with average frequency $\omega_i$, inferred from the peaks in the phonon DOS. The EP correction to the gap then reads

$$[\Delta E_g(T)]_{EP} = \sum_i A_i \cdot \left(n_B(\omega_i, T) + \frac{1}{2}\right), \tag{6}$$

where $n_B$ again stands for the Bose-Einstein occupation factor of the effective mode with frequency $\omega_i$. The analytical expression for the EP term is simply obtained by taking the first derivative of the Bose-Einstein occupation number over temperature as:

$$\left[\frac{\partial E_g}{\partial T}\right]_{EP} = \sum_i A_i \cdot \frac{\partial n_B(\omega_i, T)}{\partial T}$$
$$= \sum_i \frac{A_i}{4T} \cdot \frac{\hbar\omega_i}{k_B T} \cdot \frac{1}{sinh^2\left(\frac{\hbar\omega_i}{2k_B T}\right)}. \tag{7}$$

Here $n_B(\omega_i, T)$ is the Bose-Einstein phonon occupation factor of the $i^{th}$ oscillator. The impor-

tance of the previous equation is that allows for computing the renormalization of the gap at zero temperature, where the correction to the gap energy is given by the effect of the zero-point vibrations. Using the Einstein oscillator model the zero-point correction (ZPC) to the gap can be readily obtained by considering the limit to zero temperature of the expression in Eq. (6):

$$[\Delta E_g(0K)]_{EP} = \frac{1}{2} \cdot \sum_{i=1}^{n} A_i, \tag{8}$$

where $n = 1$ or 2 for no or finite FA content, as shown further below.

To obtain the EP term, it is convenient to perform least-squares fits using the expression derived for the first derivative of the curves gap-energy-vs-temperature. The resulting first-derivative data points (closed symbols) are shown in Figs. 3a-c of the main manuscript for x=0, 0.2 and 0.8, respectively, and in the following for the rest of FA concentrations that were studied jointly as a function of pressure and temperature. We first consider the case of the archetypal perovskite $MAPbI_3$ for which an equal footing of thermal expansion and electron-phonon interaction effects on the temperature renormalization of the gap has been obtained at room temperature.[32] To obtain the EP term (or terms) by fitting, *only* the data points within the temperature range of stability of the different crystal phases were considered, provided the TE term was also determined for the corresponding temperature range. These are the blue dots in Fig. 8 (Fig. 3a of the main manuscript) and which correspond to the TE term values resulting from Eq. (3) and tabulated in Table 1. For the cubic and tetragonal-I phases the blue dot-dashed lines indicate that the TE term is temperature independent, coinciding with the linearity of the gap temperature dependence for these phases. The contribution from electron-phonon interaction is calculated using the function of Eq. (7). However, for the orthorhombic phase the temperature dependence of the gap is clearly non-linear, which strongly speaks for a TE term that is temperature dependent. For this reason we have to pursue the measurements of the TE term at different cryogenic temperatures with the outcome tabulated in Table 1. Thus, for the fits to determine the EP term, the TE contribution will be represented by a polynomial function with fixed coefficients that inter/extrapolate the measured data points. This function is forced to be zero at the absolute zero, since $\alpha_V(T = 0) \equiv 0$. In the particular case of the or-

thorhombic phase of $MAPbI_3$ the TE term function reads: $\left(a_0 + a_1 T + a_2 T^2\right) \cdot \left( \frac{I_0 \cdot \omega_0}{4 k_B T^2 sinh^2\left(\frac{\omega_0}{2 k_B T}\right)} \right)$ with $a_0 = 0.575$ eV/K, $a_1 = -1.224 \cdot 10^{-3}$ eV/K$^2$, $a_2 = -2.97 \cdot 10^{-5}$ eV/K$^3$, $\omega_0 = 1$ meV, and $I_0 = 2.9$ eV. As demonstrated for the tetragonal-I phase of $MAPbI_3$,[32] a single Einstein oscillator is sufficient for the proper description of the EP term for the other phases too. The function of Eq. (7) together with the corresponding contribution from TE were fitted, piecewise for each phase, to the data points of Fig. 3a of the main manuscript. The resulting oscillator frequencies and amplitudes are listed in Table S8. The solid black curves and the dashed red curves in Fig. 8 (Fig. 3a) represent the resulting total rate of gap renormalization per Kelvin and the EP contributions to it, respectively.

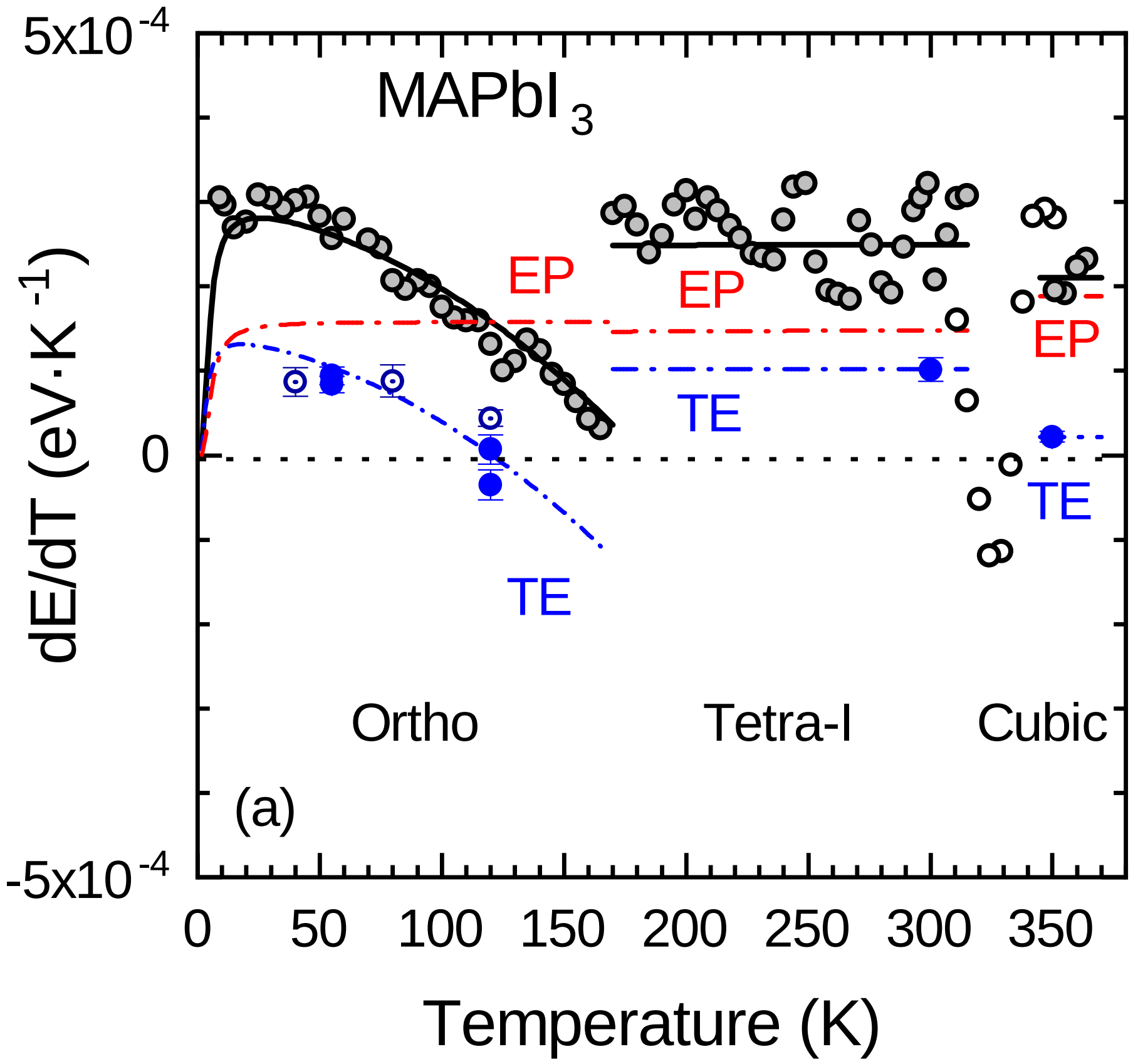


Figure S 8: Numerically calculated first derivative of the PL peak energy with respect to temperature (black-grey symbols) for $MAPbI_3$. The solid black lines represent a fit only to the data points in the temperature range of stability of the different crystal phases, corresponding to the sum of the contribution of thermal expansion (TE, blue dot-dashed curve and symbols) and electron-phonon interaction (EP, red dot-dashed curve).

Table S 8: Einstein-oscillator parameters corresponding to the effective amplitude $A_i$ and frequency $\omega_i$ describing the contribution from electron-phonon interaction $\left[\frac{\partial E_g}{\partial T}\right]_{EP}$ to the renormalization of the band gap in $MAPbI_3$. Numbers in parentheses are the error bars (uncertainty of the last digits) of the adjustable fitting parameters. Values marked with an asterisk are from Ref. [32]. Also listed are the value of the calculated zero-point corrections (ZPC) according to Eq. (8) and the temperature range in which the fit is valid.

| | **$MAPbI_3$** | | | | |
|---|---|---|---|---|---|
| Phase | Einstein osc. | $A_i$ (meV) | $\hbar\omega_i$ (meV) | ZPC (meV) | $T$ fitting range (K) |
| Cubic | Osc. ♯1 | 13(1) | 6 | | >345 |
| Tetra-I | Osc. ♯1 | 10(1)* | 5.8(6)* | | 170 - 315 |
| Ortho | Osc. ♯1 | 2.7(5) | 1.5(10) | 1 (2) | <165 |

For the orthorhombic phase of $FA_{0.2}MA_{0.8}PbI_3$ the TE term function is a third-order polynomial $a_0 + a_1 T + a_2 T^2 + a_3 T^3$ with $a_0 = 0$, $a_1 = 6.15 \cdot 10^{-6}$ eV/K$^2$, $a_2 = -6.4 \cdot 10^{-8}$ eV/K$^3$, and $a_3 = 1.33 \cdot 10^{-10}$ eV/K$^4$. Again, for the cubic and tetragonal-I phases a single Einstein oscillator is sufficient to describe the EP term, whereas for the low-temperature orthorhombic phase an anomalous electron-phonon coupling sets in that is accounted for with a second Einstein oscillator with negative amplitude. The results of the least-square fitting to the first-derivative data points of Fig. 9 (Fig. 3b of the main manuscript) corresponding to the frequency and amplitude of the different Einstein oscillators are listed in Table S9.

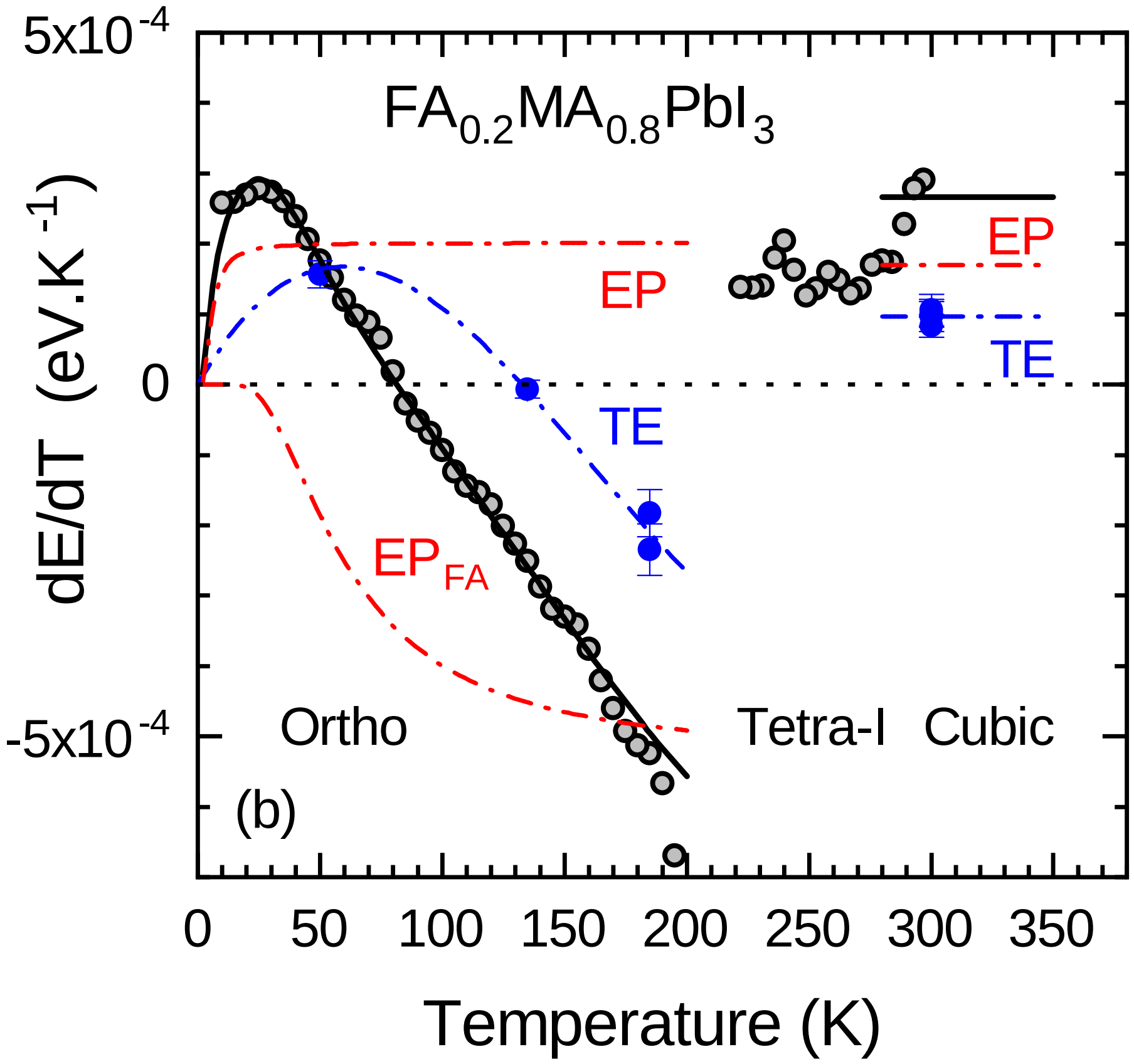


Figure S 9: Numerically calculated first derivative of the PL peak energy with respect to temperature (black-grey symbols) for $FA_{0.2}MA_{0.8}PbI_3$ mixed-cation perovskite. The solid black lines represent a fit only to the data points in the temperature range of stability of the different crystal phases, corresponding to the sum of the contribution of thermal expansion (TE, blue dot-dashed curve and symbols) and electron-phonon interaction (EP, red dot-dashed curve). The curve labeled $EP_{FA}$ represents the additional Einstein oscillator introduced to account for the anomalous electron-phonon interaction due to coupling to FA rattler modes.

Table S 9: Einstein-oscillator parameters corresponding to the effective amplitude $A_i$ and frequency $\omega_i$ describing the contribution from electron-phonon interaction $\left[\frac{\partial E_g}{\partial T}\right]_{EP}$ to the renormalization of the band gap in $FA_{0.2}MA_{0.8}PbI_3$. Numbers in parentheses are the error bars (uncertainty of the last digits) of the adjustable fitting parameters. Also listed are the value of the calculated zero-point corrections (ZPC) according to Eq. (8) and the temperature range in which the fit is valid.

| | **$FA_{0.2}MA_{0.8}PbI_3$** | | | | |
|---|---|---|---|---|---|
| Phase | Einstein osc. | $A_i$ (meV) | $\hbar\omega_i$ (meV) | ZPC (meV) | $T$ fitting range (K) |
| Cubic | Osc. ♯1 | 13(1) | 6 | | >285 |
| Tetra-I | | | | | 215 - 280 |
| Ortho | Osc. ♯1 | 3.5(5) | 1.5 | -45(15) | <170 |
| | Osc. ♯2 | -98(10) | 16(1) | | |

For the tetragonal-III phase of $FA_{0.4}MA_{0.6}PbI_3$ the TE term function is also a third-order polynomial $a_0 + a_1 T + a_2 T^2 + a_3 T^3$ with $a_0 = 0$, $a_1 = 7.34 \cdot 10^{-6}$ eV/K$^2$, $a_2 = -8.63 \cdot 10^{-8}$ eV/K$^3$, and $a_3 = 2.35 \cdot 10^{-10}$ eV/K$^4$. Again, for the cubic phase a single Einstein oscillator is sufficient to describe the EP term, whereas for the low-temperature tetragonal-III phase the second Einstein oscillator corresponding to the anomalous electron-phonon coupling also has negative amplitude. The results of the least-square fitting to the first-derivative data points of Fig. 10 corresponding to the frequency and amplitude of the different Einstein oscillators are listed in Table S10. We note that for the remaining compositions, including this one, the frequency of the FA rattler depends slightly on the temperature range considered for the least-squares fits. Typically, the smaller the range, the smaller the frequency. We thus tabulated both results but show the one corresponding to the extended $T$ range in the corresponding figures.

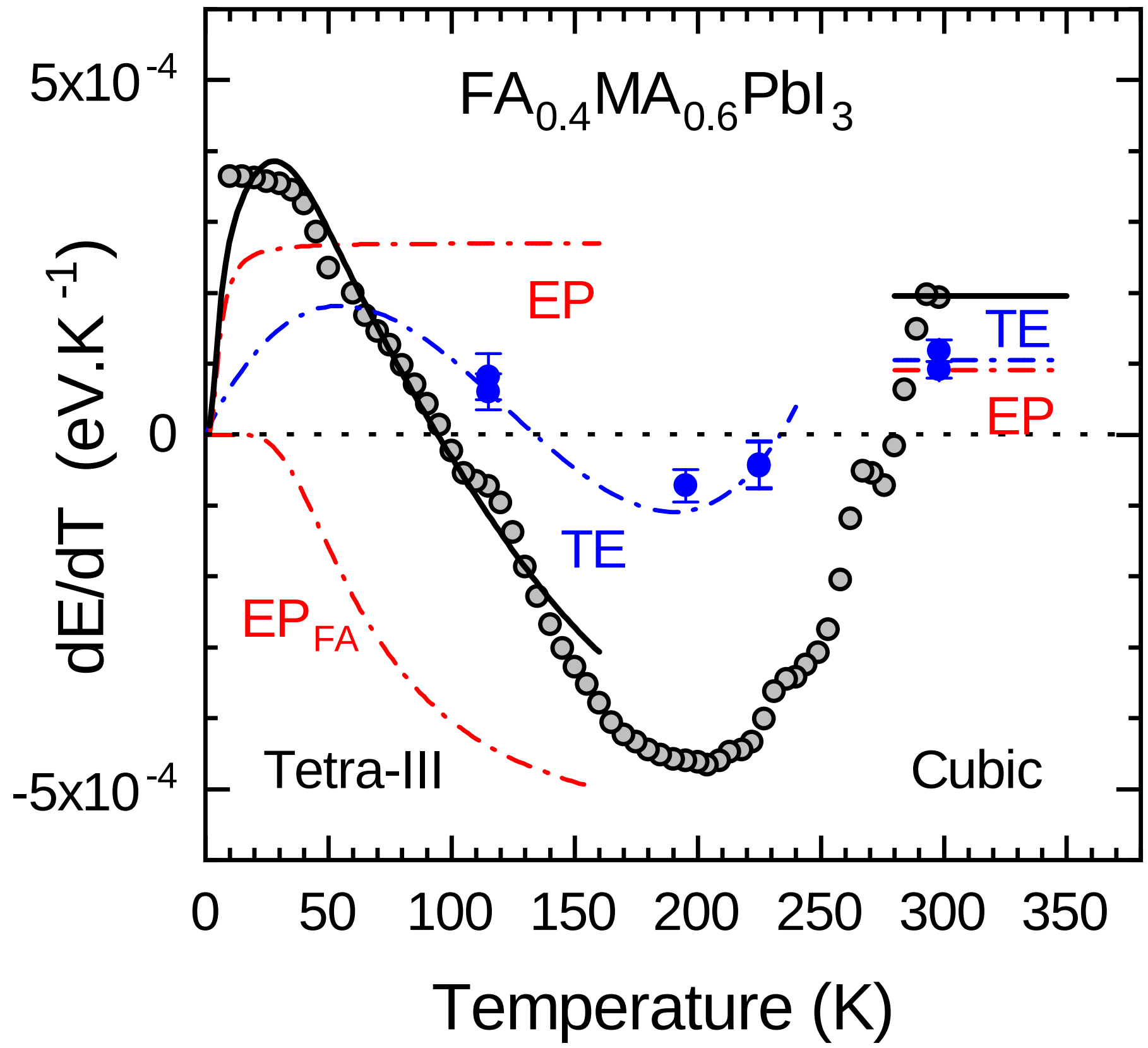


Figure S 10: Numerically calculated first derivative of the PL peak energy with respect to temperature (black-grey symbols) for $FA_{0.4}MA_{0.6}PbI_3$ mixed-cation perovskite. The solid black lines represent a fit only to the data points in the temperature range of stability of the different crystal phases, corresponding to the sum of the contribution of thermal expansion (TE, blue dot-dashed curve and symbols) and electron-phonon interaction (EP, red dot-dashed curve). The curve labeled $EP_{FA}$ represents the additional Einstein oscillator introduced to account for the anomalous electron-phonon interaction due to coupling to FA rattler modes.

Table S 10: Einstein-oscillator parameters corresponding to the effective amplitude $A_i$ and frequency $\omega_i$ describing the contribution from electron-phonon interaction $\left[\frac{\partial E_g}{\partial T}\right]_{EP}$ to the renormalization of the band gap in $FA_{0.4}MA_{0.6}PbI_3$. Numbers in parentheses are the error bars (uncertainty of the last digits) of the adjustable fitting parameters. Also listed are the value of the calculated zero-point corrections (ZPC) according to Eq. (8) and the temperature range in which the fit is valid.

| | **$FA_{0.4}MA_{0.6}PbI_3$** | | | | |
|---|---|---|---|---|---|
| Phase | Einstein osc. | $A_i$ (meV) | $\hbar\omega_i$ (meV) | ZPC (meV) | $T$ fitting range (K) |
| Cubic | Osc. ♯1 | 6(1) | 6 | | >295 |
| Tetra-III | Osc. ♯1 | 6(1) | 1.5 | -40(15) | <130 |
| | Osc. ♯2 | -85(10) | 14(1) | | |
| | Osc. ♯1 | 5(1) | 1.5 | -55(20) | <170 |
| | Osc. ♯2 | -120(10) | 18(3) | | |

For the tetragonal-III phase of $FA_{0.5}MA_{0.5}PbI_3$ the TE term function is now a fourth-order polynomial $a_0 + a_1 T + a_2 T^2 + a_3 T^3 + a_4 T^4$ with $a_0 = 0$, $a_1 = 1.22 \cdot 10^{-7}$ eV/K$^2$, $a_2 = 3.67 \cdot 10^{-8}$ eV/K$^3$, $a_3 = -4.37 \cdot 10^{-10}$, and $a_4 = 1.2 \cdot 10^{-12}$ eV/K$^5$. Again, for the cubic phase a single Einstein oscillator is sufficient to describe the EP term, whereas for the low-temperature tetragonal-III phase the second Einstein oscillator corresponding to the anomalous electron-phonon coupling also has negative amplitude. The results of the least-square fitting to the first-derivative data points of Fig. 11 corresponding to the frequency and amplitude of the different Einstein oscillators are listed in Table S11.

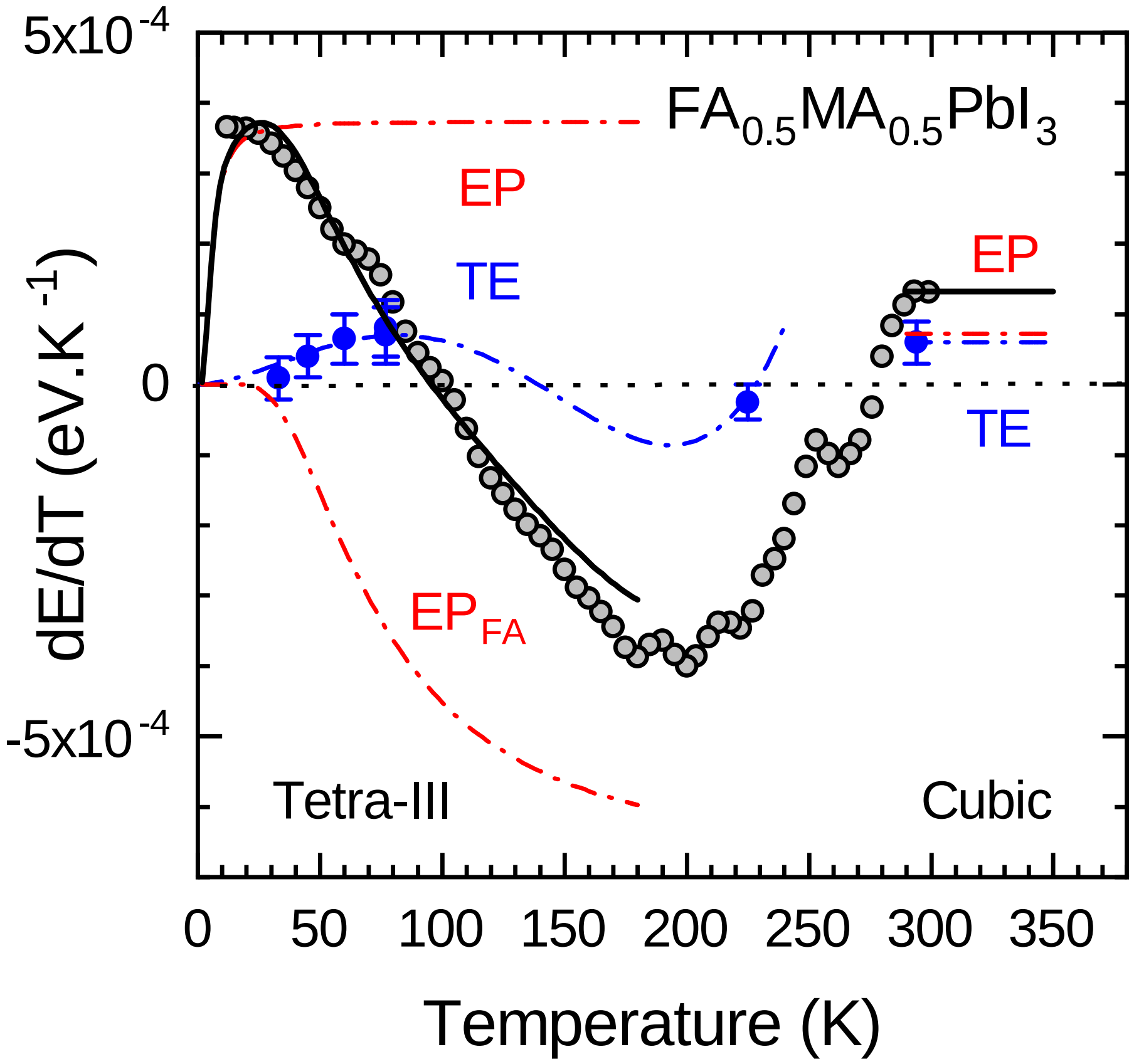


Figure S 11: Numerically calculated first derivative of the PL peak energy with respect to temperature (black-grey symbols) for $FA_{0.5}MA_{0.5}PbI_3$ mixed-cation perovskite. The solid black lines represent a fit only to the data points in the temperature range of stability of the different crystal phases, corresponding to the sum of the contribution of thermal expansion (TE, blue dot-dashed curve and symbols) and electron-phonon interaction (EP, red dot-dashed curve). The curve labeled $EP_{FA}$ represents the additional Einstein oscillator introduced to account for the anomalous electron-phonon interaction due to coupling to FA rattler modes.

Table S 11: Einstein-oscillator parameters corresponding to the effective amplitude $A_i$ and frequency $\omega_i$ describing the contribution from electron-phonon interaction $\left[\frac{\partial E_g}{\partial T}\right]_{EP}$ to the renormalization of the band gap in $FA_{0.5}MA_{0.5}PbI_3$. Numbers in parentheses are the error bars (uncertainty of the last digits) of the adjustable fitting parameters. Also listed are the value of the calculated zero-point corrections (ZPC) according to Eq. (8) and the temperature range in which the fit is valid.

| | **$FA_{0.5}MA_{0.5}PbI_3$** | | | | |
|---|---|---|---|---|---|
| Phase | Einstein osc. | $A_i$ (meV) | $\hbar\omega_i$ (meV) | ZPC (meV) | $T$ fitting range (K) |
| Cubic | Osc. ♯1 | 5(1) | 6 | | >290 |
| Tetra-III | Osc. ♯1 | 7(1) | 1.5 | -65(15) | <120 |
| | Osc. ♯2 | -135(10) | 18(2) | | |
| | Osc. ♯1 | 6(1) | 1.5 | -75(20) | <180 |
| | Osc. ♯2 | -150(10) | 19(3) | | |

For the tetragonal-III phase of $FA_{0.6}MA_{0.4}PbI_3$ the TE term function is a third-order polynomial $a_0 + a_1 T + a_2 T^2 + a_3 T^3$ with $a_0 = 0$, $a_1 = 3.82 \cdot 10^{-6}$ eV/K$^2$, $a_2 = -4.45 \cdot 10^{-8}$ eV/K$^3$, and $a_3 = 1.2 \cdot 10^{-10}$. Again, for the cubic phase a single Einstein oscillator is sufficient to describe the EP term, whereas for the low-temperature tetragonal-III phase the second Einstein oscillator corresponding to the anomalous electron-phonon coupling also has negative amplitude. The results of the least-square fitting to the first-derivative data points of Fig. 12 corresponding to the frequency and amplitude of the different Einstein oscillators are listed in Table S12.

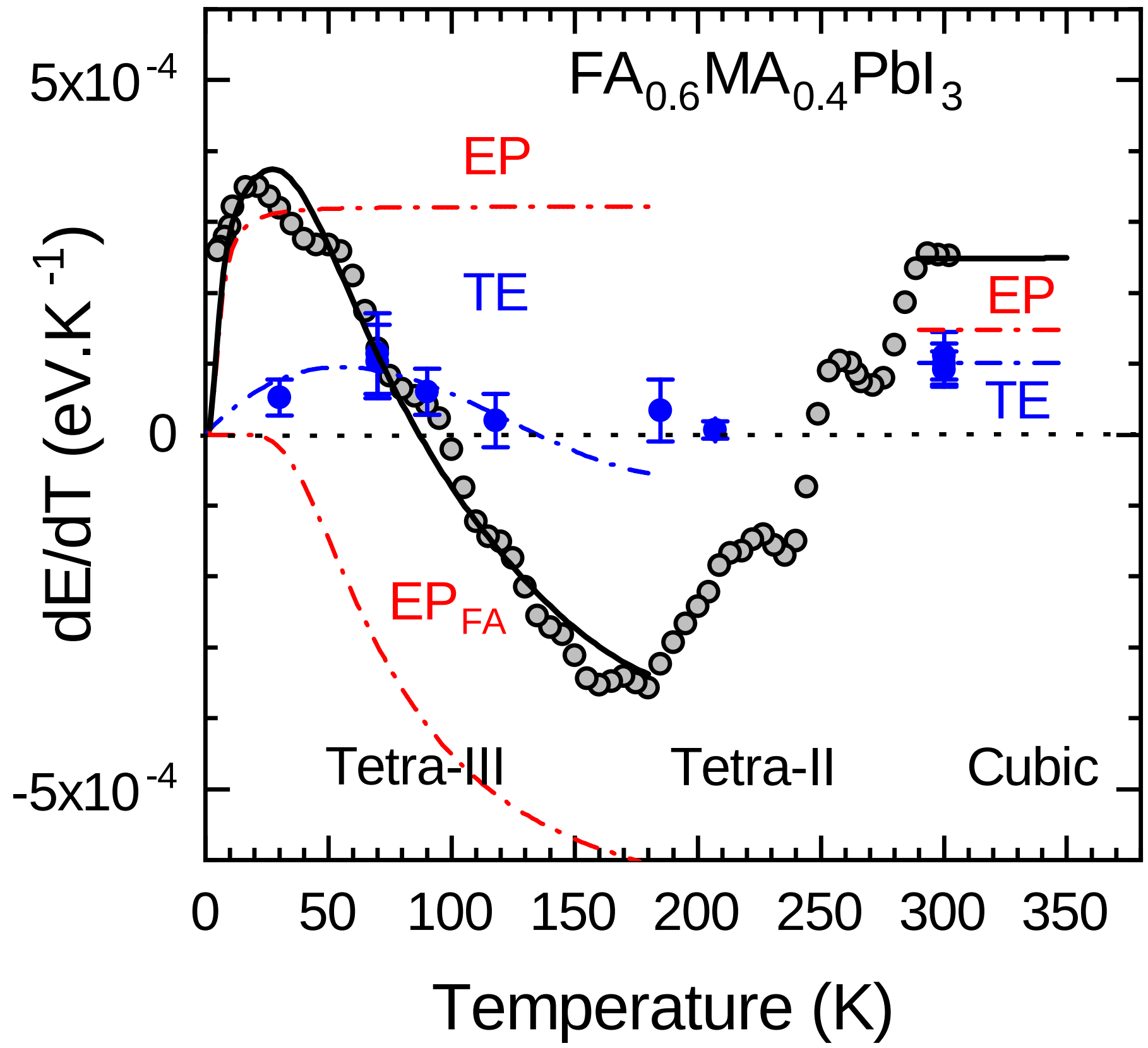


Figure S 12: Numerically calculated first derivative of the PL peak energy with respect to temperature (black-grey symbols) for $FA_{0.6}MA_{0.4}PbI_3$ mixed-cation perovskite. The solid black lines represent a fit only to the data points in the temperature range of stability of the different crystal phases, corresponding to the sum of the contribution of thermal expansion (TE, blue dot-dashed curve and symbols) and electron-phonon interaction (EP, red dot-dashed curve). The curve labeled $EP_{FA}$ represents the additional Einstein oscillator introduced to account for the anomalous electron-phonon interaction due to coupling to FA rattler modes.

Table S 12: Einstein-oscillator parameters corresponding to the effective amplitude $A_i$ and frequency $\omega_i$ describing the contribution from electron-phonon interaction $\left[\frac{\partial E_g}{\partial T}\right]_{EP}$ to the renormalization of the band gap in $FA_{0.6}MA_{0.4}PbI_3$. Numbers in parentheses are the error bars (uncertainty of the last digits) of the adjustable fitting parameters. Also listed are the value of the calculated zero-point corrections (ZPC) according to Eq. (8) and the temperature range in which the fit is valid.

| | $\mathbf{FA_{0.6}MA_{0.4}PbI_3}$ | | | | |
|---|---|---|---|---|---|
| Phase | Einstein osc. | $A_i$ (meV) | $\hbar\omega_i$ (meV) | ZPC (meV) | $T$ fitting range (K) |
| Cubic | Osc. ♯1 | 10(1) | 6 | | >290 |
| Tetra-II | | | | | 200 - 250 |
| Tetra-III | Osc. ♯1 | 6(1) | 1.5 | -60(15) | <125 |
| | Osc. ♯2 | -125(10) | 18(3) | | |
| | Osc. ♯1 | 6(1) | 1.5 | -75(20) | <180 |
| | Osc. ♯2 | -160(10) | 20(3) | | |

For the tetragonal-III phase of $FA_{0.8}MA_{0.2}PbI_3$ the TE term function is a third-order polynomial $a_0 + a_1 T + a_2 T^2 + a_3 T^3$ with $a_0 = 0$, $a_1 = 1.04 \cdot 10^{-5}$ eV/K$^2$, $a_2 = -1.29 \cdot 10^{-7}$ eV/K$^3$, and $a_3 = 3.7 \cdot 10^{-10}$. Again, for the cubic and tetragonal-II phases a single Einstein oscillator is sufficient to describe the EP term, whereas for the low-temperature tetragonal-III phase the second Einstein oscillator corresponding to the anomalous electron-phonon coupling also has negative amplitude. The results of the least-square fitting to the first-derivative data points of Fig. 13 (Fig. 3c of the main manuscript) corresponding to the frequency and amplitude of the different Einstein oscillators are listed in Table S13.

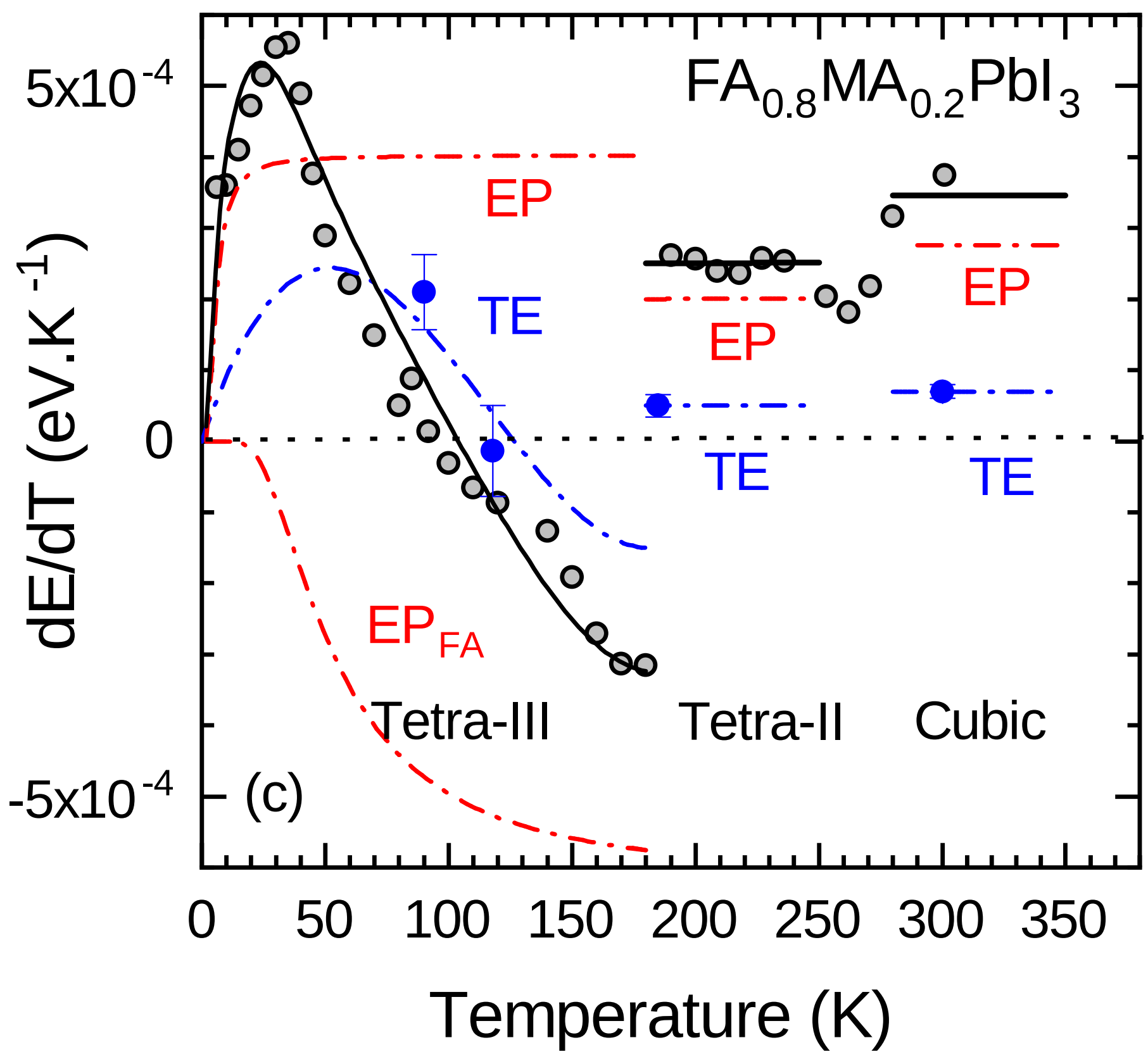


Figure S 13: Numerically calculated first derivative of the PL peak energy with respect to temperature (black-grey symbols) for $FA_{0.8}MA_{0.2}PbI_3$ mixed-cation perovskite. The solid black lines represent a fit only to the data points in the temperature range of stability of the different crystal phases, corresponding to the sum of the contribution of thermal expansion (TE, blue dot-dashed curve and symbols) and electron-phonon interaction (EP, red dot-dashed curve). The curve labeled $EP_{FA}$ represents the additional Einstein oscillator introduced to account for the anomalous electron-phonon interaction due to coupling to FA rattler modes.

Table S 13: Einstein-oscillator parameters corresponding to the effective amplitude $A_i$ and frequency $\omega_i$ describing the contribution from electron-phonon interaction $\left[\frac{\partial E_g}{\partial T}\right]_{EP}$ to the renormalization of the band gap in $FA_{0.8}MA_{0.2}PbI_3$. Numbers in parentheses are the error bars (uncertainty of the last digits) of the adjustable fitting parameters. Also listed are the value of the calculated zero-point corrections (ZPC) according to Eq. (8) and the temperature range in which the fit is valid.

| | **$FA_{0.8}MA_{0.2}PbI_3$** | | | | |
|---|---|---|---|---|---|
| Phase | Einstein osc. | $A_i$ (meV) | $\hbar\omega_i$ (meV) | ZPC (meV) | $T$ fitting range (K) |
| Cubic | Osc. ♯1 | 19(1) | 6 | | >290 |
| Tetra-II | Osc. ♯1 | 14(1) | 6 | | 180 - 250 |
| Tetra-III | Osc. ♯1 | 7(1) | 1.5 | -50(20) | <125 |
| | Osc. ♯2 | -105(10) | 14(2) | | |
| | Osc. ♯1 | 7(1) | 1.5 | -55(20) | <180 |
| | Osc. ♯2 | -115(10) | 15(2) | | |

For the tetragonal-III phase of $FA_{0.9}MA_{0.1}PbI_3$ the TE term function is a third-order polynomial $a_0 + a_1 T + a_2 T^2 + a_3 T^3$ with $a_0 = 0$, $a_1 = 5.51 \cdot 10^{-6}$ eV/K$^2$, $a_2 = -7.55 \cdot 10^{-8}$ eV/K$^3$, and $a_3 = 2.31 \cdot 10^{-10}$. Again, for the cubic and tetragonal-II phases a single Einstein oscillator is sufficient to describe the EP term, whereas for the low-temperature tetragonal-III phase the second Einstein oscillator corresponding to the anomalous electron-phonon coupling also has negative amplitude. The results of the least-square fitting to the first-derivative data points of Fig. 14 corresponding to the frequency and amplitude of the different Einstein oscillators are listed in Table S14.

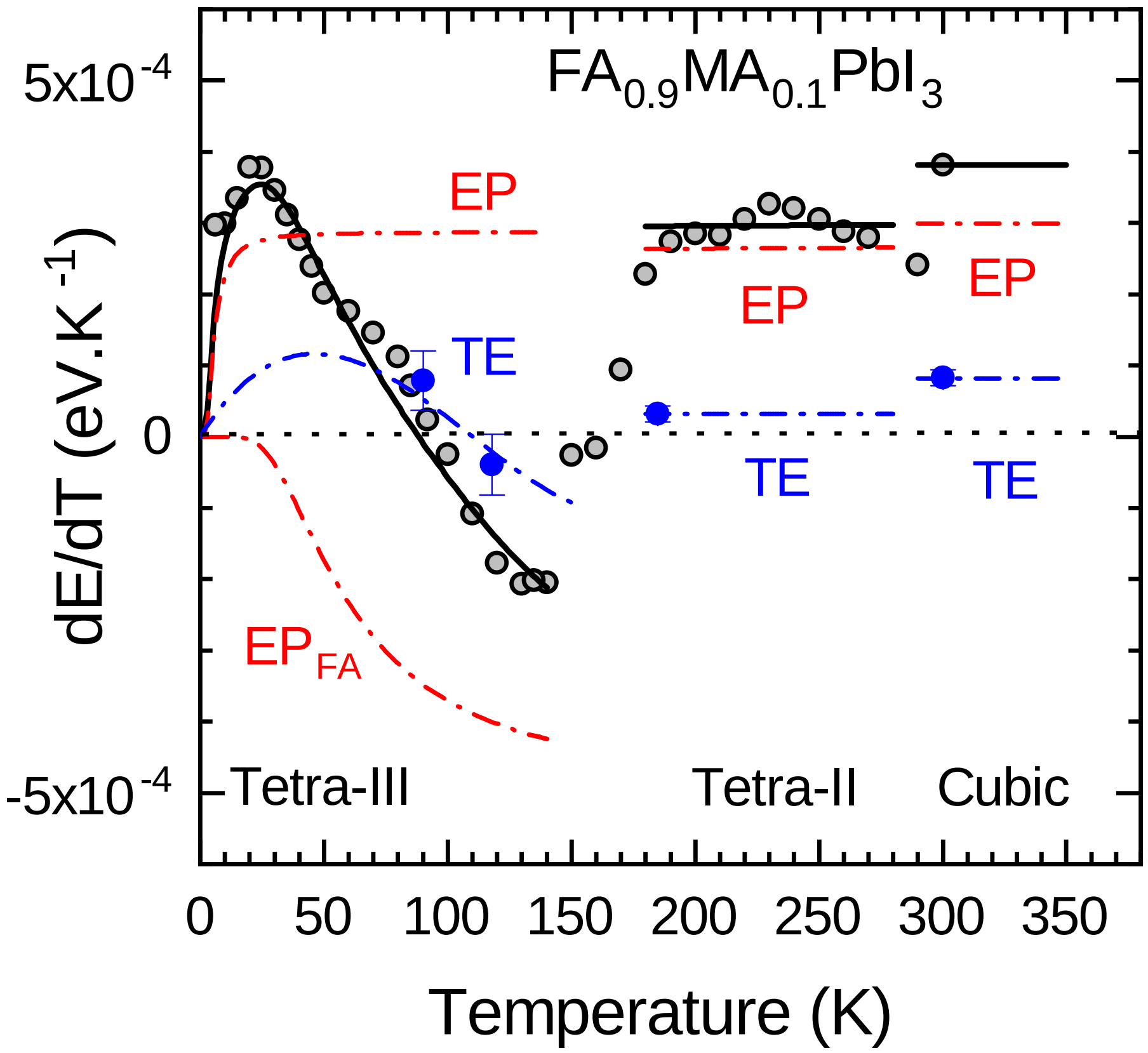


Figure S 14: Numerically calculated first derivative of the PL peak energy with respect to temperature (black-grey symbols) for $FA_{0.9}MA_{0.1}PbI_3$ mixed-cation perovskite. The solid black lines represent a fit only to the data points in the temperature range of stability of the different crystal phases, corresponding to the sum of the contribution of thermal expansion (TE, blue dot-dashed curve and symbols) and electron-phonon interaction (EP, red dot-dashed curve). The curve labeled $EP_{FA}$ represents the additional Einstein oscillator introduced to account for the anomalous electron-phonon interaction due to coupling to FA rattler modes.

Table S 14: Einstein-oscillator parameters corresponding to the effective amplitude $A_i$ and frequency $\omega_i$ describing the contribution from electron-phonon interaction $\left[\frac{\partial E_g}{\partial T}\right]_{EP}$ to the renormalization of the band gap in $FA_{0.9}MA_{0.1}PbI_3$. Numbers in parentheses are the error bars (uncertainty of the last digits) of the adjustable fitting parameters. Also listed are the value of the calculated zero-point corrections (ZPC) according to Eq. (8) and the temperature range in which the fit is valid.

| | **$FA_{0.9}MA_{0.1}PbI_3$** | | | | |
|---|---|---|---|---|---|
| Phase | Einstein osc. | $A_i$ (meV) | $\hbar\omega_i$ (meV) | ZPC (meV) | $T$ fitting range (K) |
| Cubic | Osc. ♯1 | 21(1) | 6 | | >300 |
| Tetra-II | Osc. ♯1 | 18.6(5) | 6 | | 180 - 280 |
| Tetra-III | Osc. ♯1 | 5(1) | 1.5 | -35(20) | <110 |
| | Osc. ♯2 | -75(10) | 15(2) | | |
| | Osc. ♯1 | 5(1) | 1.5 | -45(20) | <140 |
| | Osc. ♯2 | -90(10) | 16(2) | | |

According to Eq. (2), to evaluate the total gap temperature dependence (TE+EP terms) we have to calculate numerically the first derivative of the gap energies over temperature on a point-by-point basis. The black-grey symbols in Figs. S8 to S14 correspond to the first derivative data sets for the seven FA compositions measured. To obtain the EP term (or terms) for each composition by fitting, the analysis of the first-derivatives was performed piecewise within the temperature range of stability of the different phases, as indicated in the figures. The blue symbols and the blue dot-dashed curves correspond to the TE term resulting from Eq. (3) and their polynomial fits, respectively, using the parameters tabulated in Tables S1 to S7. The solid black curves in Figs. S8 to S14 represent the result of a least-squares fit to the first-derivative data points using the polynomial function describing the TE term (blue dot-dashed curve) plus the one or two Einstein-oscillator function (dashed red curve), as given by Eq. (7). The resulting amplitude and frequency of the Einstein oscillators are listed in Tables S8 to S14 for the corresponding FA contents and

plotted as a function of composition in Figs. S15a,b using closed red symbols and solid blue or blue-cyan symbols for Einstein oscillator corresponding to the normal and anomalous EP term, respectively.

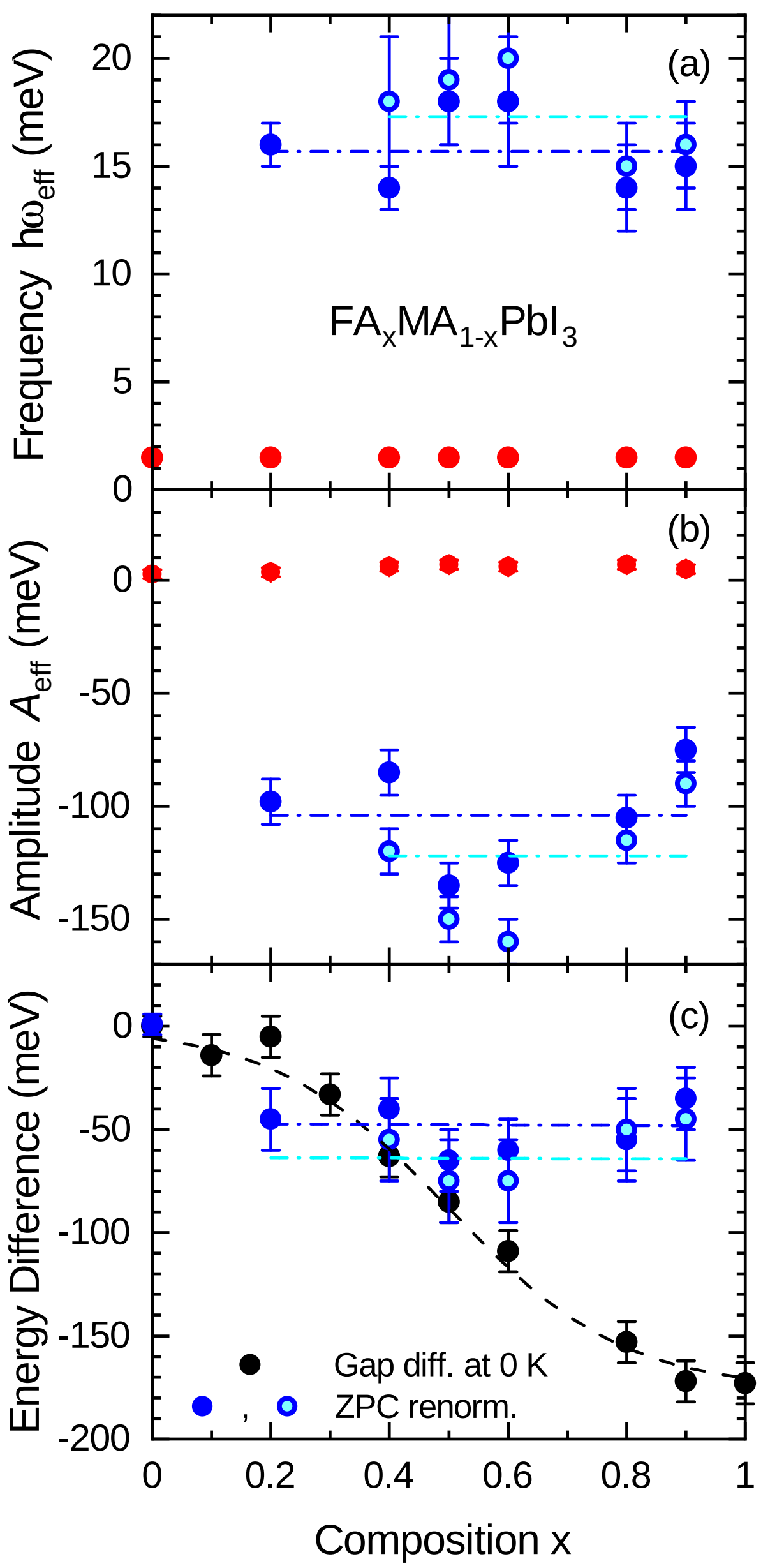


Figure S 15: **(a)** Frequency and **(b)** amplitude of the Einstein oscillator corresponding to the FA rattlers. **(c)** Differences between the gap energy extrapolated to 0 K for each composition referred to the gap of $MAPbI_3$ at 0 K plotted as a function of FA content (black symbols). Black dashed curve is a guide to the eye. Blue closed/open symbols represent the zero-point vibrations gap-corrections (ZPC) calculated using Eq. (8). Blue(cyan) dashed line represents linear fits to the closed(open) blue symbols, corresponding to the ZPC values obtained from fits performed in a shorter(wider) temperature range.

In Fig. S15c we have plotted the values of the energy difference between the gap extrapolated to 0 K for each FA composition, referred to that of $MAPbI_3$ (black symbols). The black dashed curve is a guide to the eye. Blue/blue-cyan closed symbols represent the zero-point vibrations gap corrections (ZPCs) calculated using Eq. (8). Blue(blue-cyan) dashed lines represents linear fits to the blue symbols, corresponding to the ZPC values obtained from fits performed in a shorter(wider) temperature range, as indicated in Tables S8 to S14. This comparison was made in view of the strong weight of the anomalous EP coupling in the total temperature renormalization of the gap, which made us think that it could also explain the strong composition dependence of the gap for the $FA_xMA_{1-x}PbI_3$ solid solutions. In fact, the gap energy extrapolated to 0 K (black symbols in Fig. S15c) decreases by ca. 170 meV when going from pure $MAPbI_3$ to $FAPbI_3$. We point out that this energy difference was calculated to be about 300 meV from DFT calculations including spin-orbit coupling[33] or 190 meV using the polymorphous model,[34] in excellent agreement with our experimental results. In contrast, the gap renormalization due to ZPC, though finite, it is essentially independent of FA content.

# Note ♯6: Correlation between ferroelastic-domain formation & anomalous EP coupling

In Fig. S16a, we show the temperature dependence of the gap below 250 K for three very special cases, namely for FA contents of $x$ =0, 0.1, and 0.2. The abrupt jump in gap energy signalizes the occurrence of the first-order phase transition from tetragonal (Tetra-I) to orthorhombic (Ortho). Above this transition temperature, all three compounds exhibit the same linear dependence of the gap with the typical positive slope characteristic of normal EP coupling with the phonons of the inorganic cage. However, in the Ortho phase, the temperature dependence of the gap differs in each case. For $MAPbI_3$ such dependence is non linear, though positive, and more importantly, it can be perfectly described using a single Einstein oscillator, representing the contribution from the normal EP term. In contrast, the sample with 20% FA content exhibits a pronounced bowing in

the gap temperature dependence, with a negative slope in the temperature range from 80 to 200 K. This phenomenon can only be reproduced by adding an oscillator with negative amplitude to account for the anomalous EP coupling to the FA-rattler modes.

The most interesting case is that of the 10% FA-content sample, which right after the transition into the Ortho phase exhibits a striking behavior of the gap (market with a red circle in Fig. S16a): On the one hand, there is a rigid blue-shift of five temperature data points with respect to the gap-vs-T curve at lower temperatures. On the other hand, these five points exhibit a negative slope which is exactly that of the sample with 20% FA. In Fig. S16b we reproduce the first derivative data of the sample with 10% FA content, but zooming into the temperature region of interest. We note that the precision of the measurements (scatter of the data points of the order of the symbol size) allows us to take these issues with the data points seriously. In fact, an inspection of the PL spectra of $FA_{0.1}MA_{0.9}PbI_3$ shown in Fig. S16c, corresponding to temperatures close to the Tetra-I to Ortho phase transition, tells us that the five spectra corresponding to the previously mentioned data points (wine-color curves) display a clear double-peak structure, which is indicative of the coexistence of both phases. The regions with tetragonal symmetry and larger volume exert tensile strain on the orthorhombic ones. This causes the rigid blue-shift of the gap, because its pressure coefficient is negative in this temperature range (see Table S1). The explanation for the negative temperature slope is even more enlightening. As shown in Fig. S16b, the overall tendency of the first derivative coincides very well with that of $MAPbI_3$, i.e., a gap temperature renormalization driven by thermal expansion and solely normal EP coupling (Ortho phase, blue-grey symbols in Fig. S16b). In contrast, the first derivative of these five temperature points is clearly negative (red symbols), being indicative of the contribution of the $EP_{FA}$ term with negative sign. Although the precise mechanism is still unknown, we come to the important conclusion that the strain-driven formation of ferroelastic domains is instrumental for the activation of the anomalous $EP_{FA}$ coupling term associated to the FA rattlers. These domains which are know to form in the Ortho phase after the phase transition,[35] are organized in stripes with the long tetragonal or orthorhombic axis, the preferential tilt axis, pointing at 90 deg. from stripe to stripe. Such an arrangement is the one




\begin{thebibliography}{99}

\bibitem{nrelx25a} See {\it Best Research-Cell Efficiency Chart}, source: National Renewable Energy Laboratory (NREL), Golden, Colorado, USA. www.nrel.gov/pv/cell-efficiency.html (2025).
\bibitem{linxx18a} Lin, K.; Xing, J.; Quan, L. N.; Garc\'{\i}a de Arquer, F. P.; Gong, X.; Lu, J.; Xie, L.; Zhao, W.; Zhang, D.; Yan, C.; Li, W.; Liu, X.; Lu, Y.; Kirman, J.; Sargent, E. H.; Xiong, Q.; Wei, Z. Perovskite Light-Emitting Diodes with External
Quantum Efficiency Exceeding 20 per Cent. \textit{Nature} \textbf{2018}, 562, 245-248.
\bibitem{laute85a} Lautenschlager, P.; Allen, P. B.; Cardona, M. Temperature Dependence of Band Gaps in Si and Ge. {\it Phys. Rev. B} {\bf 1985}, 31, 2163-2171.
\bibitem{gopal87a} Gopalan, S.; Lautenschlager, P.; Cardona, M. Temperature Dependence of the Shifts and Broadenings of the Critical Points in GaAs. {\it Phys. Rev. B} {\bf 1987}, 35, 5577-5584.
\bibitem{laute87a} Lautenschlager, P.; Garriga, M.; Logothetidis, S.; Cardona, M. Interband Critical Points of GaAs and their Temperature Dependence. {\it Phys. Rev. B} {\bf 1987}, 35, 9174-9189.
\bibitem{gonix98a} Go\~{n}i, A. R.; Syassen, K. Optical Properties of Semiconductors Under Pressure. {\it Semicond. Semimetals} {\bf 1998}, 54, 247-425, and references therein.
\bibitem{cardo89a} Cardona, M.; Gopalan, S. Temperature Dependence of the Band Structure of Semiconductors: Electron-Phonon Interaction. In: Doni, E.; Girlanda, R.; Parravicini, G.P.; Quattropani, A. (eds) \textit{Progress in Electron Properties of Solids. Physics and Chemistry of Materials with Low-Dimensional Structures} (Springer, Dordrecht, 1989), vol. 10, p. 51-64. https://doi.org/10.1007/978-94-009-2419-2
\bibitem{rubin21a} Rubino, A.; Francisco-L\'{o}pez, A.; Barker, A. J.; Petrozza, A.; Calvo, M. E.; Go\~{n}i, A. R.; M\'{\i}guez, H. Disentangling Electron-Phonon Coupling and Thermal Expansion Effects in the Band Gap Renormalization of Perovskite Nanocrystals. \textit{J. Phys. Chem. Lett.} \textbf{2021}, 12, 569-575.
\bibitem{goebe98a} G\"{o}bel, A.; Ruf, T.; Cardona, M.; Lin, C. T.; Wrzesinski, J.; Steube, M.; Reimann, K.; Merle, J.-C.; Joucla, M. Effects of the Isotopic Composition on the Fundamental Gap of CuCl. {\it Phys. Rev. B} {\bf 1998}, 57, 15183-15190.
\bibitem{serra02a} Serrano, J.; Schweitzer, Ch.; Lin, C. T.; Reimann, K.; Cardona, M.; Fr\"{o}hlich, D. Electron-Phonon Renormalization of the Absorption Edge of the Cuprous Halides. {\it Phys. Rev. B} {\bf 2002}, 65, 125110/1-7.
\bibitem{bhosa12a} Bhosale, J.; Ramdas, A. K.; Burger, A.; Mu\~{n}oz, A.; Romero, A. H.; Cardona, M.; Lauck, R.; Kremer, R. K. Temperature Dependence of Band Gaps in Semiconductors: Electron-Phonon Interaction. {\it Phys. Rev. B} {\bf 2012}, 86, 195208/1-10.
\bibitem{shang23a} Shang, H.; Yang, J. The Electron-Phonon Renormalization in the Electronic Structure Calculation: Fundamentals, Current Status, and Challenges. \textit{J. Chem. Phys.} \textbf{2023}, 158, 130901/1-11.
\bibitem{fasah25a} Fasahat, S.; Sch\"{a}fer, B.; Xu, K.; Fiuza-Maneiro, N.; G\'{o}mez-Gra\~{n}a, S.; Alonso, M. I.; Polavarapu, L.; Go\~{n}i, A. R. Absence of Anomalous Electron-Phonon Coupling in the Temperature Renormalization of the Gap of CsPbBr$_3$ Nanocrystals. \textit{J. Phys. Chem. C} \textbf{2025}, 129, 453-463. %\textbf{2024} https://doi.org/10.1021/acs.jpcc.4c06265. %arXiv:2409.06374 [cond-mat.mtrl-sci] (https://doi.org/10.48550/arXiv.2409.06374).
\bibitem{franc19a} Francisco L\'{o}pez, A.; Charles, B.; Weber, O. J.; Alonso, M. I.; Garriga, M.; Campoy-Quiles, M.; Weller, M. T.; Go\~{n}i, A. R. Equal Footing of Thermal Expansion and Electron-Phonon Interaction in the Temperature Dependence of Lead Halide Perovskite Band Gaps. {\it J. Phys. Chem. Lett.} {\bf 2019}, 10, 2971-2977.
\bibitem{zacha23a} Zacharias, M.; Volonakis, G.; Giustino, F.; Even, J. Anharmonic Electron-Phonon Coupling in Ultrasoft and Locally Disordered Perovskites. \textit{npj Comput. Mater.} \textbf{2023}, 9, 153/1-13.
\bibitem{zacha26a} Zacharias, M.; Volonakis, G.; Pedesseau, L.; Katan, C.; Giustino, F.; Even, J. Electron-Phonon Couplings in Polymorphous Crystals. \textit{Phys. Rev. B} \textbf{2026}, 113, L081104/1-10.
\bibitem{zacha26b} Zacharias, M.; Volonakis, G.; Pedesseau, L.; Katan, C.; Giustino, F.; Even, J. Roadmap for Electronic Structure, Anharmonicity, and Electron-Phonon Calculations in Locally Disordered Inorganic and Hybrid Halide Perovskites. \textit{Phys. Rev. B} \textbf{2026}, 113, 085118/1-45.
\bibitem{pieni23a} Pieniazek, A.; Dybala, F.; Polak, M. P.; Przypis, L.; Herman, A. P.; Kopaczek, J.; Kudrawiec, R. Bandgap Pressure Coefficient of a CH$_3$NH$_3$PbI$_3$ Thin Film Perovskite, \textit{J. Phys. Chem. Lett.} \textbf{2023}, 14, 6470-6476.
\bibitem{pieni25a} Pieniazek, A.; Dybala, F.; Przypis, L.; Polak, M. P.; Norek, M.; Kowalski, B. J.;  Kudrawiec, R. Beyond the Cubic Phase: Pressure-Induced Bandgap Modulation in a CH$_3$NH$_3$PbBr$_3$ Perovskite at Low Temperatures. \textit{Adv. Optical Mater.} \textbf{2025}, e03177/1-12. 
\bibitem{saran17a} Saran, R.; Heuer-Jungemann, A.; Kanaras, A. G.; Curry, R. J. Giant Bandgap Renormalization and Exciton-Phonon Scattering in Perovskite Nanocrystals. {\it Adv. Optical Mater.} {\bf 2017}, 5, 1700231/1-9.
\bibitem{xuxxx23a} Xu, F.; Wei, H.; Wu, Y.; Zhou, Y.; Li, J.; Cao, B. Nonmonotonic Temperature-Dependent Bandgap Change of CsPbCl$_3$ Films Induced by Optical Phonon Scattering. \textit{J. Lumin.} \textbf{2023}, 257, 119736/1-7.
\bibitem{fasah25b} Fasahat, S.; Fiuza-Maneiro, N.; Sch\"{a}fer, B.; Xu, K.; G\'{o}mez-Gra\~{n}a, S.; Alonso, M. I.; Polavarapu, L.; Go\~{n}i, A. R. Sign of the Gap Temperature Dependence in CsPb(Br,Cl)$_3$ Nanocrystals Determined by Cs-Rattler-Mediated Electron-Phonon Coupling. \textit{J. Phys. Chem. Lett.} \textbf{2025}, 16, 1134-1141.
\bibitem{lahns24a} Lahnsteiner, J.; Rang, M.; Bokdam, M. Tuning Einstein Oscillator Frequencies of Cation Rattlers: A Molecular Dynamics Study of the Lattice Thermal Conductivity of CsPbBr$_3$. \textit{J. Phys. Chem. C} \textbf{2024}, 128, 1341-1349.
\bibitem{zheng17a} Zheng, H.; Dai, J.; Duan, J.; Chen, F.; Zhu, G.; Wang, F.; Xu, C. Temperature-Dependent Photoluminescence Properties of Mixed-Cation Methylammonium-Formamidium Lead Iodide [HC(NH$_2$)$_2$]$_x$[CH$_3$NH$_3$]$_{1-x}$PbI$_3$ Perovskite Nanostructures. \textit{J. Mater. Chem. C} \textbf{2017}, 5, 12057-12061.
\bibitem{franc20a} Francisco-L\'{o}pez, A., Charles, B., Alonso, M. I., Garriga, M., Campoy-Quiles, M., Weller, M. T., Go\~{n}i, A. R. Phase Diagram of Methylammonium/Formamidinium Lead Iodide Perovskite Solid Solutions from Temperature-Dependent Photoluminescence and Raman Spectroscopies. {\it J. Phys. Chem. C} {\bf 2020}, \textit{124}, 3448-3458.
\bibitem{haine25a} Hainer, T.; Fransson, E.; Dutta, S.; Wiktor, J.; Erhart, P. A Morphotropic Phase Boundary in MA$_{1-x}$FA$_x$PbI$_3$: Linking Structure, Dynamics and Electronic Properties. \textit{Nat. Commun.} \textbf{2025}, 16, 8775/1-9.
\bibitem{simen24a} Simenas, M.; Gagor, A.; Banys, J.; Maczka, M. Phase Transitions and Dynamics in Mixed Three- and Low-Dimensional Lead Halide Perovskites. \textit{Chem. Rev.} \textbf{2024}, 124, 2281-2326.
\bibitem{weber16a} Weber, O. J.; Charles, B. L.; Weller, M. T. Phase Behaviour and Composition in the Formamidinium-Methylammonium hybrid Lead Iodide Perovskite Solid Solution. \textit{J. Mater. Chem. A} \textbf{2016}, 4, 15375-15382.
\bibitem{weber18a} Weber, O. J.; Ghosh, D.; Gaines, S.; Henry, P. F.; Walker, A. B.; Islam, M. S.; Weller, M. T. Phase Behavior and Polymorphism of Formamidinium Lead Iodide. \textit{Chem. Mater.} \textbf{2018}, 30, 3768-3778.
\bibitem{mohan19a} Mohanty, A.; Swain, D.; Govinda, S.; Row, T. N. G.; Sarma, D. D. Phase Diagram and Dielectric Properties of MA$_{1-x}$FA$_x$PbI$_3$. \textit{ACS Energy Lett.} \textbf{2019}, 4, 2045-2051.
\bibitem{dutta25a} Dutta, S.; Fransson, E.; Hainer, T.; Gallant, B. M.; Kubicki, D. J.; Erhart, P.; Wiktor, J. Revealing the Low-Temperature Phase of FAPbI$_3$ Using a Machine-Learned Potential. \textit{J. Am. Chem. Soc.} \textbf{2025}, 147, 37019-37029.
\bibitem{konto20a} Kontos, A. G.; Manolis, G. K.; Kaltzoglou, A.; Palles, D.; Kamitsos, E. I.; Kanatzidis, M. G.; Falaras. P. Halogen-NH$_2^+$ Interaction, Temperature-Induced Phase Transition, and Ordering in (NH$_2$CHNH$_2$)PbX$_3$ (X = Cl, Br, I) Hybrid Perovskites. \textit{J. Phys. Chem. C} \textbf{2020}, 124, 8479-8487.
\bibitem{weixx23a} Wei, Y.; Volosniev, A. G.; Lorenc, D.; Zhumekenov, A. A.; Bakr, O. M.; Lemeshko, M.; Alpichshev, Z. Bond Polarizability as a Probe of Local Crystal Fields in Hybrid Lead-Halide Perovskites. \textit{J. Phys. Chem. Lett.} \textbf{2023}, 14, 6309-6314.
\bibitem{xuxxx23b} Xu, K.; P\'{e}rez-Fidalgo, L.; Charles, B. L.; Weller, M. T.; Alonso, M. I.; Go\~{n}i, A. R. Using Pressure to Unravel the Structure\textemdash Dynamic-Disorder Relationship in Metal Halide Perovskites. \textit{Sci. Rep.} \textbf{2023}, 13, 9300/1-12.
\bibitem{gonix26a} Go\~{n}i, A. R. Phonon Interactions in Metal Halide Perovskites Elucidated by Raman Scattering. To be published in \textit{Appl. Phys. Rev.} \textbf{2026}. 
\bibitem{welle15a} Weller, M. T.; Weber, O. J.; Henry, P. F.; Di Pumpo, A. M.; Hansen, T. C. Complete Structure and Cation Orientation in the Perovskite Photovoltaic Methylammonium Lead Iodide between 100 and 352 K. {\it Chem. Commun.} {\bf 2015}, 51, 4180-4183.
\bibitem{welle15b} Weller, M. T.; Weber, O. J.; Frost, J. M.; Walsh, A. Cubic Perovskite Structure of Black Formamidinium Lead Iodide, $\alpha$-[HC(NH$_2$)$_2$]PbI$_3$, at 298 K. {\it J. Phys. Chem. Lett.} {\bf 2015}, 6, 3209-3212.
\bibitem{saida15a} Saidaminov, M. I.; Abdelhady, A. L.; Murali, B.; Alarousu, E.; Burlakov, V. M.; Peng, W.; Dursun, I.; Wang, L.; He, Y.; Maculan, G.; {\it et al.} %Goriely, A.; Wu, T.; Mohammed, O. F.; Bakr, O.M.
    High-Quality Bulk Hybrid Perovskite Single Crystals within Minutes by Inverse Temperature Crystallization. {\it Nature Commun.} {\bf 2015}, 6, 7586.
\bibitem{leguy16a} Leguy, A. M. A.; Go\~{n}i, A. R.; Frost, J. M.; Skelton, J.; Brivio, F.; Rodr\'{\i}guez-Mart\'{\i}nez, X.; Weber, O. J.; Pallipurath, A.; Alonso, M. I.; Campoy-Quiles, M.; Weller, M. T.; Nelson, J.; Walsh, A.; Barnes, P. R. F. Dynamic Disorder, Phonon Lifetimes, and the Assignment of Modes to the Vibrational Spectra of Methylammonium Lead Halide Perovskites. {\it Phys. Chem. Chem. Phys.} {\bf 2016}, 18, 27051-27066.
\bibitem{galco17a} Galkowski, K.; Mitioglu, A. A.; Surrente, A.; Yang, Z.; Maude, D. K.; Kossaki, P.; Eperon, G. E.; Wang, J. T.-W.; Snaith, H. J.; Plochocka, P.; Nicholas, R. J. Spatially Resolved Studies of the Phases and Morphology of Methylammonium and Formamidinium Lead Tri-Halide Perovskites. {\it Nanoscale} {\bf 2017}, 9, 3222-3230.
\bibitem{hanse24a} Hansen, K. R.; Colton, J. S.; Whittaker-Brooks, L. Measuring the Exciton Binding Energy: Learning from a Decade of Measurements on Halide Perovskites and Transition Metal Dichalcogenides. \textit{Adv. Optical Mater.} \textbf{2024}, 12, 2301659/1-136.
\bibitem{alons19a} Alonso, M. I.; Charles, B.; Francisco-L\'{o}pez, A.; Garriga, M.; Weller, M. T.; Go\~{n}i, A. R. Spectroscopic Ellipsometry Study of FA$_x$MA$_{1-x}$PbI$_3$ Hybrid Perovskite Single Crystals. {\it J. Vac. Sci. Technol. B} {\bf 2019}, 37, 062901/1-7.
\bibitem{orans18a} Oranskaia, A.; Yin, J.; Bakr, O. M.; Br\'{e}das, J.-L.; Mohammed, O. F. Halogen Migration in Hybrid Perovskites: The Organic Cation Matters. {\it J. Phys. Chem. Lett.} {\bf 2018}, 9, 5474-5480.
\bibitem{meggi18a} Meggiolaro, D.; De Angelis, F. First-Principle Modeling of Defects in Lead Halide Perovskites: Best Practices and Open Issues. {\it ACS Energy Lett.} {\bf 2018}, 3, 2206-2222.
\bibitem{franc21a} Francisco-L\'{o}pez, A.; Charles, B.; Alonso, M. I.; Garriga, M.; Weller, M. T.; Go\~{n}i, A. R. Photoluminescence of Bound-Exciton Complexes and Assignment to Shallow Defects in Methylammonium/Formamidinium Lead Iodide Mixed Crystals. {it Adv. Optical Mater.} \textbf{2021}, 2001969/1-9.
\bibitem{charl17a} Charles, B.; Dillon, J.; Weber, O. J.; Islam, M. S.; Weller, M. T. Understanding the Stability of Mixed A-Cation Lead Iodide Perovskites. {\it J. Mater. Chem.A} {\bf 2017}, 5, 22495-22499.
\bibitem{perez23a} P\'{e}rez-Fidalgo, L.; Xu, K.; Charles, B. L.; Henry, P. F.; Weller, M. T.; Alonso, M. I.; Go\~{n}i, A. R. Anomalous Electron-Phonon Coupling in Cesium-Substituted Methylammonium Lead Iodide Perovskites. \textit{J. Phys. Chem. C} \textbf{2023}, 127, 22817-22826.
\bibitem{franc18a} Francisco L\'{o}pez, A.; Charles, B.; Weber, O. J.; Alonso, M. I.; Garriga, M.; Campoy-Quiles, M.; Weller, M. T.; Go\~{n}i, A. R. Pressure-Induced Locking of Methylammonium Cations Versus Amorphization in Hybrid Lead Iodide Perovskites. {\it J. Phys. Chem. C} {\bf 2018}, 122, 22073-22082.
\bibitem{posto17a} Postorino, P.; Malavasi, L. Pressure-induced effects in organic-inorganic hybrid perovskites. {\it J. Phys. Chem. Lett.} {\bf 2017}, 8, 2613-2622.
\bibitem{jacob15a} Jacobsson, T. J.; Schwan, L. J.; Ottosson, M.; Hagfeldt, A.; Edvinsson, T. Determination of Thermal Expansion Coefficients and Locating the Temperature-Induced Phase Transition in Methylammonium Lead Perovskites Using X-ray Diffraction, \textit{Inorg. Chem.} \textbf{2015}, 54, 10678-10685.
\bibitem{whitf17a} Whitfield, P. S.; Herron, N.; Guise, W. E.; Page, K.; Cheng, Y. Q.; Milas, I.; Crawford, M. K. Corrigendum: Structures, Phase Transitions and Tricritical Behavior of the Hybrid Perovskite Methyl Ammonium Lead Iodide, \textit{Sci. Rep.} \textbf{2017}, 7, 42831.
\bibitem{whitf16a} Whitfield, P. S.; Herron, N.; Guise, W. E.; Page, K.; Cheng, Y. Q.; Milas, I.; Crawford, M. K. Structures, Phase Transitions and Tricritical Behavior of the Hybrid Perovskite Methyl Ammonium Lead Iodide, \textit{Sci. Rep.} \textbf{2016}, 6, 35685/1-15.
\bibitem{fabin16a} Fabini, D. H.; Stoumpos, C. C.; Laurita, G.; Kaltzoglou, A.; Kontos, A. G.; Falaras, P.; Kanatzidis, M. G.; Seshadri, R. Reentrant Structural and Optical Properties and Large Positive Thermal Expansion in Perovskite Formamidinium Lead Iodide, \textit{Angew. Chem.} \textbf{2016}, 128, 15618-15622.
\bibitem{jaffe16a} Jaffe, A.; Lin, Y.; Beavers, C. M.; Voss, J.; Mao, W. L.; Karunadasa, H. I. High-Pressure Single-Crystal Structures of 3D Lead-Halide Hybrid Perovskites and Pressure Effects on their Electronic and Optical Properties. {\it ACS Cent. Sci.} {\bf 2016}, 2, 201-209.
\bibitem{szafr16a} Szafra\'{n}ski, M.; Katrusiak, A. Mechanism of Pressure-Induced Phase Transitions, Amorphization, and Absorption-Edge Shift in Photovoltaic Methylammonium Lead Iodide. {\it J. Phys. Chem. Lett.} {\bf 2016}, 7, 3458-3466.
\bibitem{corde19a} Cordero, F.; Craciun, F.; Trequattrini, F.; Generosi, A.; Paci, B.; Paoletti, A. M.; Pennesi, G. Stability of Cubic FAPbI$_3$ from X-ray Diffraction, Anelastic, and Dielectric Measurements, \textit{J. Phys. Chem. Lett.} \textbf{2019}, 10, 2463-2469.
\bibitem{zhaox20a} Zhao, X.-G.; Dalpian, G. M.; Wang, Z.; Zunger, A. Polymorphous Nature of Cubic Halide Perovskites, \textit{Phys. Rev. B} \textbf{2020}, 101, 155137/1-19.
\bibitem{evenx16a} Even, J.; Carignano, M.; Katan, C. Molecular Disorder and Translation/Rotation Coupling in the Plastic Crystal Phase of Hybrid Perovskites, \textit{Nanoscale} \textbf{2016}, 8, 6222-6236.
\bibitem{zacha20a} Zacharias, M.; Giustino, F. Theory of the Special Displacement Method for Electronic Structure Calculations at Finite Temperature, \textit{Phys. Rev. Res.} \textbf{2020}, 2, 013357/1-24.
\bibitem{kongx16a} Kong, L.; Liu, G.; Gong, J.; Hu, Q.; Schaller, R. D.; Dera, P.; Zhang, D.; Liu, Z.; Yang, W.; Tang, Y.; Wang, C.; Wei, S.-H.; Xu, T.; Mao, H.-K. Simultaneous Band-Gap Narrowing and Carrier-Lifetime Prolongation of Organic-Inorganic Trihalide Perovskites. {\it PNAS} {\bf 2016}, 113, 8910-8915.
\bibitem{senno21a} Senno, M.; Tinte, S. Mixed Formamidinium-Methylammonium Lead Iodide Perovskite from First-Principles: Hydrogen Bonding Impact on the Electronic Properties. \textit{Phys. Chem. Chem. Phys.} \textbf{2021}, 23, 7376-7385.
\bibitem{bergm92a} Thomas, L. K. in \textit{Bergmann-Schaefer, Lehrbuch der Experimentalphysik-Festk\"orper}, ed. by Raith, W. (Walter de Gruyter, Berlin, \textbf{1992}), Vol. 6, p. 364. ISBN: 3-11-012605-2.
\bibitem{simen21a} Simenas, M.; Balciunas, S.; Svirskas, S.; Kinka, M.; Ptak, M.; Kalendra, V.; Gagor, A.; Szewczyk, D.; Sieradzki, A.; Grigalaitis, R.; Walsh, A.; Maczka, M.; Banys, J. Phase Diagram and Cation Dynamics of Mixed MA$_{1-x}$FA$_x$PbBr$_3$ Hybrid Perovskites. \textit{Chem. Mater.} \textbf{2021}, 33, 5926-5934. 
\bibitem{ambro22a} Ambrosio, F.; De Angelis, F.; Go\~{n}i, A. R. The Ferroelectric-Ferroelastic Debate about Metal Halide Perovskites. {\it J. Phys. Chem. Lett.} {\bf 2022}, \textit{13}, 7731-7740.
\bibitem{kimxx17a} Kim, H.; Hunger, J.; C\'{a}novas, E.; Karakus, M.; Mics, Z.; Grechko, M.; Turchinovich, D.; Parekh, S. H.; Bonn, M. Direct Observation of Mode-Specific Phonon-Band Gap Coupling in Methylammonium Lead Halide Perovskites. {\it Nature Commun.} {\bf 2017}, 8, 687/1-9.
\bibitem{guoxx20a} Guo, P.; Xia, Y.; Gong, J.; Cao, D. H.; Li, X.; Li, X.; Zhang, Q.; Stoumpos, C. C.; Kirschner, M. S.; Wen, H.; Prakapenka, V. B.; Ketterson, J. B.; Martinson, A. B. F.; Xu, T.; Kanatzidis, M. G.; Chan, M. K. Y.; Schaller, R. D. Direct Observation of Bandgap Oscillations Induced by Optical Phonons in Hybrid Lead Iodide Perovskites. \textit{Adv. Funct. Mater.} \textbf{2020}, 30, 1907982/1-8.

\end{thebibliography}
\end{document}